\documentclass[11pt]{article}
\usepackage{a4wide}
\usepackage{amsmath, amssymb, amscd, amsthm, amsfonts}
\usepackage[utf8]{inputenc}
\usepackage{graphicx}
\usepackage{hyperref}
\usepackage{cite}
\usepackage[dvipsnames]{xcolor}

\newcommand{\e}{{\cal E}}
\newcommand{\s}{{\cal S}}

\newcommand{\aop}[3][]{\hat{#2}_{#1 \bf #3}} 
\newcommand{\vt}[1]{{\small \bf #1}} 
\newcommand{\dpr}[2]{{\bf #1} \cdot {\bf #2}} 
\newcommand{\Mp}{M_{\rm Pl}} 
\newcommand{\ket}[1]{\left| #1 \right\rangle} 
\newcommand{\bra}[1]{\left\langle #1 \right|} 
\newcommand{\Tr}{{\rm Tr}}

\begin{document}
{\renewcommand{\thefootnote}{\fnsymbol{footnote}}
		
\begin{center}
{\LARGE Quantum corrections to the primordial tensor spectrum:\\
Open EFTs \& Markovian decoupling of UV modes} 
\vspace{1.5em}

Suddhasattwa Brahma$^{1}$\footnote{e-mail address: {\tt suddhasattwa.brahma@gmail.com}}, 
Arjun Berera$^{1}$\footnote{e-mail address: {\tt ab@ed.ac.uk}}, and 
Jaime Calder\'on-Figueroa$^{1}$\footnote{e-mail address: {\tt jaime.calderon@ed.ac.uk}}
\\
\vspace{1.5em}
$^1$Higgs Centre for Theoretical Physics, School of Physics and Astronomy,\\ University of Edinburgh, Edinburgh EH9 3FD, UK\\

\vspace{1.5em}
\end{center}
}
	
\setcounter{footnote}{0}

\begin{abstract}
	\noindent Perturbative quantum corrections to primordial power spectra are important for testing the robustness and the regime of validity of inflation as an effective field theory. Although this has been done extensively for the density power spectrum (and, to some extent, for the tensor spectrum) using loop corrections,  we do so in an open quantum system approach to the problem. Specifically, we calculate the first order corrections to the primordial gravitational wave spectrum due to (cubic) tensor interactions alone. We show that our results match expectations from standard loop corrections only in the \textit{strict Markovian limit}, and therefore, establish a systematic way to relax this approximation in the future, as is generally necessary for gravitational systems.
\end{abstract}

\section{Introduction}
As with any successful theory of nature, it is important to question the realm of validity of inflation as an effective field theory (EFT) on curved spacetimes \cite{Weinberg:2005vy, Weinberg:2006ac}. Inflation is generally considered as the relevant theory to describe long-wavelength degrees of freedom (dofs) as originating from quantum fluctuations in the very early universe. However, a natural question often posed is how far back can we trust inflation as an EFT of quantum fields on a quasi-de Sitter (dS) background. This typically goes under the name of the 'trans-Planckian problem' of inflationary cosmology \cite{Martin:2000xs, Brandenberger:2000wr}. Simply put, classical inhomogeneities observed at late times, if traced back long enough, could have arisen from trans-Planckian modes, a regime in which the predictions of inflation cannot be trusted anymore \cite{Brandenberger:2021kjo}. Instead of going into the debate of understanding the pros and cons of whether this is indeed a physical problem (see, \textit{e.g.}  \cite{Brandenberger:2021pzy, Berera:2020dvn, Kaloper:2018zgi, Burgess:2020nec} for recent perspectives), a safe statement is that going beyond linear interactions between the cosmological fluctuations should be a pointer towards understanding the depth of this problem. In other words, if this was to be a physical problem, then quantum corrections to the linear two-point function (\textit{i.e.} the primordial power spectrum) would lead to large $\mathcal{O}(1)$ corrections to the free theory calculation since the interactions would become strongly coupled at the Planck scale ($\Mp$) \cite{Shiu:2005si}. 

The reader might pause at this point and say that we already know that loop corrections to the inflationary power spectrum are small \cite{Senatore:2009cf, Pimentel:2012tw}, and therefore, this unambiguously shows that there is no trans-Planckian problem of inflationary cosmology. However, the main point of this article will be to show that the standard arguments regarding the nonlinear quantum effects to the two-point function is correct to the leading order, and yet there are many subtleties which have to do with the limits of validity of the approximations which go into standard loop corrections. To identify these caveats, let us begin by first spelling out that we are interested in the EFT of the super-Hubble modes during inflation which later re-enter the Hubble horizon to become late-time inhomogeneities \cite{Burgess:2014eoa, Boyanovsky:2015jen, Boyanovsky:2015tba, Burgess:2015ajz, Hollowood:2017bil}. In other words, we are choosing to divide our full Hilbert space into system and environment dofs (as is always done in physics) -- either in the form of `heavy' \textit{vs.} `light' modes, `slow' \textit{vs.} `fast' variables, `high' \textit{vs.} `low' energies and so on. And even then, what is remarkable about our choice is that this immediately tells us that the usual formulation of organizing an EFT in terms of energies -- the standard \textit{Wilsonian effective action} -- will not be suitable for us. 

Let us unpack the last assertion in a bit more detail. The physical wavelength of primordial perturbations are stretched due to gravitational red-shifting over cosmological expansion. In the case of inflation, if we demarcate super-Hubble modes as the `system' dofs\footnote{This is conceptually a big leap from standard QFT in dS space which often goes unnoticed due to the omnipresent status of inflation in early-universe cosmology. For our setup, the modes which `exit' the horizon reenters after some time since subsequent cosmological evolution is dominated by radiation and matter epochs, unlike what would have been the case for pure dS space.}, then the Hilbert space corresponding to these keeps growing with time. In these types of scenarios, the system can gain or lose energy to the bath (or environment), and thus, the system undergoes non-unitary evolution \cite{Shandera:2017qkg, Brahma:2020zpk, EE_bounce}. In other words, there is no `energy conservation' law implying that the heavy modes which are \textit{integrated out} can reappear later on. This forbids us to write down an effective Wilsonian action for the low-energy dofs in terms of operators defined only in the low-energy Hilbert space.  All of this points towards adopting a more general EFT formulation for cosmological perturbations.

In fact, it is often the case that in theories with horizons, it is better to use an open quantum systems approach \cite{Kaplanek:2019dqu, Kaplanek:2019vzj}. In open EFTs, one only has access to the system modes while tracing out over the environment modes. The crucial difference with standard Wilsonian integrating out is that the unobserved sector is not demarcated on the basis of energy of the modes. Rather, we will consider the physical wavelength to be the discrepancy parameter: The super-Hubble modes during inflation will be our system and we will trace out over all sub-Hubble dofs, allowing for the possibility that some of the environment dofs will become part of the system Hilbert space due to gravitational red-shifting. In this way, we use the natural Hubble horizon provided to us by the dynamics of the system as the demarcating scale -- one that is of extreme importance since it has been long argued that modes classicalize only after crossing the Hubble horizon \cite{Nelson:2016kjm, Gong:2019yyz, Martin:2018lin, Martin:2018zbe}.

When adopting such a new approach to inflation as an open EFT, several fresh perspectives are bound to come up. The one we will seek to address in this work is to what approximation are we allowed to ignore the quantum entanglement between different scales \cite{Brahma:2020zpk}, and the associated dissipation, when calculating correlation functions? Functionally, in such an open EFT, one calculates the reduced density matrix of the system modes and then uses this density matrix to calculate correlation functions between operators living solely in the system Hilbert space. If indeed we were dealing with a standard QFT in flat space where our well-trusted energy conservation law was valid, then it can be shown that the low-energy density matrix, related to the vacuum state of the field theory, is equivalent to using the Wilsonian effective action \cite{Balasubramanian:2011wt}. However, this is precisely the point  --- such an effective action is not available in the case of inflation! To further stress this point, recall that gravitational interactions has this unique feature that they are universal and cannot be `turned off' at will. Thus, unlike the quintessential example of collider experiments in high-energy physics, the interactions between the short and long-wavelength modes always remain active and thus one \textit{cannot} assume the short-wavelength dofs to remain in their ground state. As shown in \cite{Agon:2014uxa}, this invalidates the standard Wilsonian argument and necessitates the generalization to Feynman-Vernon influence functionals. 

In this paper, however, we will not use the path integral formulation but rather stick to the canonical version of using master equations which determine the evolution of the reduced density matrix. What we aim to do is calculate the first order quantum corrections to the power spectrum due to cubic interactions. so how does this differ from standard loop corrections? Since we have argued that the usual Wilsonian EFT formulation does not work anymore, we cannot then use the same tools to calculate radiative corrections in the standard manner. More physically, the open EFT we are introducing to the game takes into account the dissipation happening between the system and environment degrees of freedom. Of course, we still need to make approximations in order to do explicit calculation. Although abandoning energy hierarchies, one often uses the time scales associated with the system and environment. If the quantity of interest happens over a long enough time-scale such that the system can relax to equilibrium before being perturbed by the environment, then one can systematically ignore the entanglement between the system modes and the environment (even though the interaction is never turned off). We will show that when solving the master equation assuming such a `Markovian approximation', the results, to leading order, indeed match with what one gets from the one-loop corrections in standard QFT in a quasi-dS background.

This is the main finding of our work -- although the leading-order quantum corrections due to one-loop effects (using the well-known `in-in' formalism) match with those when using an open EFT approach, it does so \textit{only under the Markovian approximation}. The validity of this assumption is not clearly understood for gravitational systems, and this gives us a systematic way to find corrections to this approximation in an open EFT setup. If using the standard `in-in' formulation to calculate loop corrections to primordial correlation functions, one would be blind to such effects which arise due to the quantum entanglement between long and short dofs. And, of course, if trans-Planckian modes are ever to play a role in the quantum corrections, it will precisely be due to this fact that they cannot be safely `disentangled' or `renormalized' from the observable modes of interest. In this paper, we do not answer whether this is indeed the case, or not, for inflation; rather, we open a new direction for doing such calculations and show that the well-known results match only when imposing the `Markovian' approximation.

In the next subsection, we outline the technical details of the problem at hand and give a quick overview of the existing literature on the topic. In section \ref{Sec:2}, we introduce the interaction Hamiltonian and set up our master equation in section \ref{Sec:3}. Section \ref{sec:PS} elaborates the solution for the primordial tensor power spectrum and first order quantum corrections to it. We conclude in section \ref{sec:CN} by discussing the significance of our result and ways to generalize it in future. Most of the technical details have been relegated to the various appendices.

\subsection{Quantum corrections to the power spectrum: Review of existing literature}
Specifically, we are interested in calculating the first order quantum correction to the primordial tensor power spectrum due to cubic interactions between the tensor modes alone. Comparing with the standard EFT language, our result should be compared with the one-loop correction to the tensor power spectrum due to cubic tensor interactions. The reason for studying this specific type of corrections is that this arises from the purely gravitational part of the action and does not involve the slow-roll parameters for the dominant terms of interest. The cubic interaction is purely due to the non-linearity of gravity. In fact, this is the system of interest for most models of inflation which do not introduce any new rank-two massless tensor. To be clear, there can obviously be other one-loop effects to the tensor power spectrum which are bigger than the one we are considering; however, for any model of inflation, the interaction between the primordial gravity waves is going to be of this form (to leading order) as long as one has general relativity as the effective theory in mind. Contrast this with the case of scalar (or density) perturbations, whose leading order (cubic) interactions can easily be modified from the single-clock slow roll case by introducing new features in the potential or by adding higher derivative terms in the action. 

In order to do a like-for-like comparison, we need contrast our result \textit{vis a vis} the one loop correction to the gravitational wave power spectrum due to (cubic) pure tensor mode interactions. However, to the best of our knowledge, this calculation has not been done in the past. Nevertheless, there are similar calculations which have been done which can help guide us in our quest. Once again, we are looking for the result of such one loop corrections to the tensor power spectrum \textit{only} to compare with our result and our calculation for computing the first-order quantum corrections due to the sub-Hubble (environment) tensor modes do not rely on this. Our sole motivation for this is to check how far standard loop corrections are able to capture the correct physics of inflation (when there is no longer any time-translation symmetry) when compared with the full machinery of open EFTs. In the process, we fill a gap in our current understanding of how quantum corrections from purely tensor perturbations affect the primordial tensor power spectrum.

To this end, we note that the case of loop corrections to primordial power spectra has been studied for more than a decade now. The initial motivation was indeed to test the limits of validity of inflation as an EFT and if such quantum corrections, howsoever tiny, can open new directions in understanding the physics of inflation (or, more generally, that of dS space) \cite{Weinberg:2005vy}. First, the case of loop corrections to the scalar power spectrum were studied in detail in \cite{Senatore:2009cf, Pimentel:2012tw}. It was found that loop corrections do not spoil the scale invariance of the spectrum \cite{Senatore:2012ya}. Rather, the loop corrections to the inflationary power spectrum has a logarithmic dependence on the renormalization scale. Since then, further studies have only strengthened this result when considering different types of matter fields when coupled to the inflaton. In a similar way, the one loop correction to the tensor power spectrum has been calculated due to coupling to free fermions (both massive and massless) as well as due to coupling to other types of matter components such as isocurvature fields \cite{Adshead:2009cb, Frob:2012ui, Tan:2019czo}. For our purposes, the case closest to us was studied in \cite{Tan:2019czo} where the author considered the one-loop correction of the tensor power spectrum due to a massless minimally-coupled scalar field. We shall return to a detailed comparison of their findings with our result in the discussion section, but let us summarize the main conclusions of loop corrections people have reached over the years with regards to inflationary perturbations. 

Although it was initially thought that the scale invariance of the tree-level power spectrum will be spoiled due to a logarithmic dependence on the comoving momentum coming from one-loop effects \cite{Weinberg:2005vy, Sloth:2006az, Seery:2007we, Adshead:2009cb, Kahya:2010xh}, the common consensus seems to be that such terms must cancel amongst themselves, and we would find no such enhancement to the power spectrum due to one-loop corrections \cite{Pimentel:2012tw, Tanaka:2013caa, Tan:2019czo, Frob:2012ui}. We do not provide any new or alternative viewpoint on this and, in any case, we are not interested in computing a loop correction in this work. Rather, our goal is to provide a fresh perspective on what should be the physical way to compute quantum corrections for an open EFT such as inflation. In the beginning, as will be demonstrated later on, the calculations in the open EFT approach does seem to suggest that the perturbative quantum corrections would end up involving the comoving momentum of the subhorizon modes. However, at the very end, we do find that the final result is actually scale-invariant, as has been long speculated starting from the seminal work on \cite{Senatore:2009cf}. What we add to this discussion, due to the nature of the new techniques employed in this work, is that how \textit{only under certain simplifying assumptions} does one reach this familiar result, and how this opens up the possibility of going beyond such approximations to calculate next-to-leading order corrections for perturbative quantum effects during inflation. We will also show that this assumption of (time-local) Markovian behaviour is a good one, but not an exact one, and one needs to relax this for exploring further.

\section{Action and mode expansions for tensor perturbations}\label{Sec:2}
We systematically introduce the quadratic action and the cubic interaction terms involving tensor modes alone and, in the process, set up our notation.

\subsection{The quadratic action}
We consider that the cosmological background is described by the flat, FLRW metric 
\begin{eqnarray}
	ds^2 = - dt^2 + a^2(t) dx^2 = a^2(\tau) \left(-d\tau^2 + dx^2\right)\,,
\end{eqnarray}
where $t$ and $\tau$ are cosmic and conformal times respectively. For inflation, we have the usual relation: $a(\tau) \sim 1/H\tau$, where $H$ is the (approximately constant) Hubble scale during inflation. To consider the action for cosmological perturbations, we can go to the Hamiltonian (ADM) formalism: 
\begin{eqnarray}
	ds^2 = - N^2 dt^2 + q_{ij} \left(dx^i + N^i dt\right) \left(dx^j + N^j dt\right)\,.
\end{eqnarray}
Since we are only interested in interactions purely between primordial gravitational waves, it is sufficient to consider the metric on the spatial hypersurface $q_{ij} = a^2\left(\delta_{ij} + h_{ij}\right)$, where we have denoted tensor modes by $h_{ij}$ and have not considered any scalar (or vector) perturbations.

As is well known, the above prescription leads to a quadratic action for tensor perturbations which is given by \cite{Maldacena:2002vr} (primes denote derivatives with respect to conformal time $\tau$)
\begin{equation}
    {\cal S}_2^{(t)} = \frac{\Mp^2}{8} \int d\tau\ d^3x\ a^2 \left[(h_{ij}')^2 - (\partial h_{ij})^2 \right]\,,
\end{equation}
where the tensor field, and its corresponding polarization tensors, are given by
\begin{equation}
    h_{ij} = \sum_{\alpha = 1}^{2} h_{\alpha} e_{ij}^{\alpha}\,, \qquad e_{ij}^{\alpha}e_{ij}^{\alpha'} = 2 \delta^{\alpha \alpha'}\,.
\end{equation}
It is customary to work with the $+$ and $\times$ polarizations, and throughout this work, we will mostly follow the conventions of \cite{Gong:2019yyz}. In general, these tensors can be built from two vectors perpendicular to the momentum $\vt{k}$ in the following way, 
\begin{equation}
	e^{+}_{ij} = (\vt{e}_1)_i  (\vt{e}_1)_j -  (\vt{e}_2)_i   (\vt{e}_2)_j, \qquad e^{\times}_{ij} = (\vt{e}_1)_i  (\vt{e}_2)_j +  (\vt{e}_2)_i   (\vt{e}_1)_j,
\end{equation}
where $\vt{e}_1 \times \vt{e}_2 = \hat{\vt{k}}$. Eqs. \eqref{polten1}-\eqref{polten3} in Appendix \ref{ap:DK} show the explicit form of these tensors for three particular choices of $\hat{\vt{k}}$.

The linear equations of motion simplify significantly if we introduce the canonical variable (and also conveniently brings this problem closer to the case of scalar perturbations)
\begin{equation}
    v_{\alpha} := \frac{a \Mp}{\sqrt{2}}h_{\alpha}\,,
\end{equation}
such that the quadratic action becomes
\begin{equation}\label{S2}
    {\cal S}_2^{(t)} = \frac{1}{2} \sum_{\alpha = +,\times} \int d\tau\ d^3x\ \left[ (v_{\alpha}')^2 - (\partial v_{\alpha})^2 + \frac{a''}{a}v_{\alpha}^2 \right]\,.
\end{equation}
It reproduces the well-known fact that, at linear level, the gravitational wave action looks like two copies of that for a minimally-coupled scalar field. Consequently, the equation of motion takes the form
\begin{equation}
    v_k'' + \left(k^2 - \frac{a''}{a}\right)v_k = 0\,,
\end{equation}
where, assuming the Bunch-Davies vacuum, the solution for the mode functions are
\begin{equation}
    v_k (\tau) = \frac{e^{-ik\tau}}{\sqrt{2k}}\left(1 - \frac{i}{k\tau}\right)\,.
\end{equation}

In this way we can write the tensor field in the Heisenberg picture as
\begin{equation}\label{ladder_op}
    \hat{h}_{ij} (\tau, \vt{x}) = \int \frac{d^3 k}{(2\pi)^3} \frac{\sqrt{2}}{a \Mp} \sum_{\alpha} \left[ v_k (\tau) e_{ij}^{\alpha} (\vt{k}) \aop{c}{k}^{\alpha} + v_k^* (\tau) e_{ij}^{\alpha} (-\vt{k}) \aop{c}{-k}^{\alpha \dagger} \right]e^{i \dpr{k}{x}}\,.
\end{equation}
This expansion can be similarly found through the rotation 
\begin{equation}
    \hat{h}_{ij} (\tau, \vt{x}) = U_{0}^{\dagger} (\tau;\tau_0) \hat{h}_{ij} (\vt{x}) U_0 (\tau;\tau_0)\,,
\end{equation}
where $ U_0 (\tau;\tau_0)$ is the evolution operator corresponding to the free Hamiltonian which can be schematically expressed as
\begin{eqnarray}\label{Evol_Free}
	 U_0 (\tau;\tau_0) := T\, \exp{\left(-i \int_{\tau_0}^\tau d\tau \,H_2^{(t)}\right)}\,,
\end{eqnarray}
where the quadratic Hamiltonian can easily be derived from the action \eqref{S2} and $T$ denotes time-ordering. Note that the quadratic Hamiltonian is not time-independent due to a time-dependent mass (squeezing) term as discussed below. $\hat{h}_{ij} (\vt{x})$ is the Schr\"odinger operator given by 
\begin{equation}
    \hat{h}_{ij} (\vt{x}) = \int \frac{d^3 k}{(2\pi)^3} \frac{1}{\sqrt{k}a \Mp} \left[e_{ij}^{\alpha} (\vt{k}) \aop{c}{k}^{\alpha} + e_{ij}^{\alpha}(-\vt{k}) \aop{c}{-\vt{k}}^{\alpha \dagger}\right] e^{i\dpr{k}{x}}\,.
\end{equation}
The quadratic Hamiltonian, corresponding to the action \eqref{S2}, has two pieces -- the standard Hamiltonian for free scalar fields in flat space and a term which couples $\vt{k}$ with $-\vt{k}$, known as the squeezing operator. These squeezing interaction can be thought of as a time-dependent mass term which result in a gravitational pump sourcing zero-momentum correlated pairs. The unitary evolution operators depend on the so-called squeezing parameters and are defined in the same fashion as for scalar perturbations \cite{Albrecht:1992kf}. Although all of these things are well-known, we repeat them to emphasize that the evolution generated by the quadratic Hamiltonian do not lead to any mode-coupling, and therefore, they do not contribute to the entanglement between modes of different wavelengths. Only cubic interactions, which we introduce in the next subsection, will be responsible for such couplings.

\subsection{Cubic interaction Hamiltonian}
The cubic action which describes all the self-interactions of  primordial tensor modes is given by \cite{Prokopec:2010be, Gong:2019yyz}
\begin{align}
    {\cal S}_3^{(t)} = \int d\tau d^3 x\ a^2 \Mp^2 & \left[ -\frac{1}{2} h_{ij}h_{jk}'h_{ki}' - 2 {\cal H} h_{ij}h_{jk}h_{ki}' + 2 \left(1-\frac{\epsilon}{3}\right) {\cal H}^2 h_{ij}h_{jk}h_{ki} \right. \nonumber \\ 
    & \left. + h_{ij} \left(\frac{1}{4}h_{kl,i}h_{kl,j} + \frac{1}{2}h_{ik,l}h_{jl,k} - \frac{3}{2}h_{ik,l}h_{jk,l}\right) \right]\,,
\end{align}
where we have introduced the slow-roll parameter $\epsilon := -\dot{H}/H^2$, and ${\cal H} := a'/a = a H$, where $H$ denotes the (approximately) constant Hubble parameter. From the above action, we can derive the interaction Hamiltonian as 
\begin{align}\label{HIf}
    H_{\rm int} = -a^2 \Mp^2 \int d^3 x  & \left[ -\frac{1}{2} h_{ij}h_{jk}'h_{ki}' - 2 {\cal H} h_{ij}h_{jk}h_{ki}' + 2 \left(1-\frac{\epsilon}{3}\right) {\cal H}^2 h_{ij}h_{jk}h_{ki} \right. \nonumber \\ 
    & \left. + h_{ij} \left(\frac{1}{4}h_{kl,i}h_{kl,j} + \frac{1}{2}h_{ik,l}h_{jl,k} - \frac{3}{2}h_{ik,l}h_{jk,l}\right) \right]\,.
\end{align}
One thing which is important to point out here is that there are terms in the above cubic action which are independent of any `so-called' slow-roll parameters. This is so because there are tensor mode perturbations even for pure dS whereas scalar perturbations can only appear for a quasi-dS background.\footnote{It is indeed possible to think of scalar modes as Goldstone bosons which appear due to the breaking of time-translation symmetry in an EFT for inflationary perturbations \cite{Cheung:2007st}.} This fact implies some simplifications regarding gauge issues when dealing with pertubative quantum corrections to the tensor power spectrum, when considering tensor interactions alone. 

Next, we Fourier expand the operator above and go to the interaction picture, as it will simplify much of the computations. The standard way to pass over from the Schr\"odinger picture to the interaction one is given by $\mathcal{O}_{\rm I} :=  U_0^\dagger (\tau;\tau_0)\, \mathcal{O}_{\rm Schr}\,  U_0 (\tau;\tau_0)$, where $U_0 (\tau;\tau_0)$ is the unitary evolution operator corresponding to the quadratic Hamiltonian introduced in Eqn. \eqref{Evol_Free}. Following these steps, we obtain 
\begin{align}\label{HI}
    H_{\rm int, I} := V_{I}(\tau) = \int_{\Delta_k} \sum_{\alpha_1, \alpha_2, \alpha_3} & \left\{ h_0(\tau, \vt{k_1}, \vt{k_2}, \vt{k_3}) \aop{c}{k_1}^{\alpha_1} \aop{c}{k_2}^{\alpha_2} \aop{c}{k_3}^{\alpha_3} (\tau_0) + h_1(\tau,-\vt{k_1},\vt{k_2},\vt{k_3}) \left[ \aop{c}{-k_1}^{\alpha_1 \dagger} \aop{c}{k_2}^{\alpha_2} \aop{c}{k_3}^{\alpha_3} (\tau_0) \right. \right. \nonumber \\
    & \left.\left.  + \aop{c}{k_3}^{\alpha_3} \aop{c}{-k_1}^{\alpha_1 \dagger} \aop{c}{k_2}^{\alpha_2} (\tau_0) + \aop{c}{k_2}^{\alpha_2} \aop{c}{k_3}^{\alpha_3} \aop{c}{-k_1}^{\alpha_1 \dagger} (\tau_0) \right] \right\} + {\rm h.c.}\,,
\end{align}
where we have defined
\begin{equation}
    \int_{\Delta_k} := \int \frac{d^3 k_1}{(2\pi)^3} \frac{d^3 k_2}{(2\pi)^3} \frac{d^3 k_3}{(2\pi)^3} (2\pi)^3 \delta^{(3)} (\vt{k_1} + \vt{k_2} + \vt{k_3})\,,
\end{equation}
and
\begin{equation}
    h_0 (\tau,\vt{k_1}, \vt{k_2}, \vt{k_3}) \approx -\frac{4\sqrt{2}}{a \Mp} \left(1-\frac{\epsilon}{3}\right) {\cal H}^2 e_{ij}^{\alpha_1} (\vt{k_1})e_{jk}^{\alpha_2}(\vt{k_2})e_{ki}^{\alpha_3}(\vt{k_3}) v_{k_1}^{\alpha_1}(\tau) v_{k_2 }^{\alpha_2}(\tau) v_{k_3}^{\alpha_3}(\tau)\;,
\end{equation}

\begin{equation}
    h_1 (\tau,\vt{k_1}, \vt{k_2}, \vt{k_3}) \approx -\frac{4 \sqrt{2}}{a \Mp} \left(1-\frac{\epsilon}{3}\right) {\cal H}^2 e_{ij}^{\alpha_1} (\vt{k_1})e_{jk}^{\alpha_2}(\vt{k_2})e_{ki}^{\alpha_3}(\vt{k_3}) [v_{k_1}^{\alpha_1}(\tau)]^* v_{k_2 }^{\alpha_2}(\tau) v_{k_3}^{\alpha_3}(\tau)\;.
\end{equation}
The tensor fields are expanded in terms of the creation and annihilation operators introduced in Eqn. \eqref{ladder_op}, so we have all the ingredients ready for setting up the master equation for the super-Hubble modes.

Before moving on, note the following two properties. Firstly, the interaction Hamiltonian, as written in Eqn. \eqref{HI}, is manifestly hermitian. Furthermore, we only consider the third term on the r.h.s. of Eqn. \eqref{HIf} since the others are subdominant. The reason is twofold. First, the chosen term is the only one with no time or spatial derivatives. Thus, requirements like slow-roll or background homogeneity pose no constraints on it. Moreover, momenta factors coming from the spatial derivatives become more relevant in the UV, and so it is expected that their contribution when involving IR modes ($\s$) will be suppressed in comparison to terms with no such derivatives.

\section{Master Equation and Lindblad Operators}\label{Sec:3}
Before writing down the master equation governing the evolution of the reduced density matrix corresponding to the super-Hubble modes, let us introduce some necessary notation. First, unless stated otherwise (through a subscript), every operator is written in the Schr\"odinger picture. Likewise, the subscripts $\e$ and $\s$ indicate that operators are acting on the environment or system Hilbert space, respectively. If no subscript is pointed out, then the operator acts on the entire Hilbert space. Notice that for any operator which spans the entire Hilbert space, we have the decomposition
\begin{equation}
    \hat{A} = \sum_i \hat{A}_{\e}^i \otimes \hat{A}_{\s}^i\;,
\end{equation}
where the sum is over any possible combination of system and environment modes associated to the Fourier expansion of the operator. Naturally, this expansion is going to be very messy, but the resulting expressions can be reduced by relabeling momenta and using other symmetries (related to permutations of the indices), similar to what was done in Eqn. (\ref{HI}).  

A full derivation of the master equation is shown in Appendix \ref{Ame}, where we find that
\begin{align}\label{meq0}
    \frac{d \rho_{\rm red}(\tau)}{d\tau} = & -i \left[H_{0,\s}, \rho_{\rm red}^{(0)} + \rho_{\rm red}^{(1)} + \rho_{\rm red}^{(2)} \right] -i \left[V_{\rm eff1} + V_{\rm eff2}, \rho_{\s}(\tau) \right] \nonumber \\
    & - \frac{1}{2} \sum_i \left[ L_1^{\dagger}L_2 \rho_\s (\tau) + \rho_\s (\tau) L_2^{\dagger} L_1 - 2 L_1 \rho_\s (\tau) L_2^{\dagger} + (L_1 \leftrightarrow L_2 ) \right]\;,
\end{align}
where 
\begin{equation}
    \rho_{\rm red}^{(0)} (\tau) := \rho_\s (\tau) = U_{0,\s}\ket{\s_0}\bra{\s_0}U_{0,\s}^{\dagger} = \ket{\s(\tau)}\bra{\s(\tau)}\,,
\end{equation}
\begin{equation}
    V_{\rm eff1} := \left\langle \e (\tau_0) \left| U_{0,\e}^{\dagger} (\tau; \tau_0) V_{\s}\, U_{0, \e}(\tau;\tau_0) \right| \e (\tau_0) \right\rangle\,, \quad V_{\rm eff2} := -\frac{i}{2} \sum_i \left(L_1^{\dagger}L_2 - L_2^{\dagger}L_1\right)\,,
\end{equation}
\begin{equation}
    L_1 := \Big\langle \e_i \Big| V_{\s}\, U_{0,\e}(\tau;\tau_0)\Big| \e (\tau_0) \Big\rangle\,, \quad    L_2 := \left\langle \e_i \left| \int_{\tau_0}^{\tau} d\tau' V_I (\tau' - \tau) U_{0,\e}(\tau;\tau_0) \right| \e(\tau_0) \right\rangle\,,
\end{equation}
where $V:= V_{\rm Schr}$ denotes an interaction term in the Schr\"odinger picture and $V_I (\tau'-\tau) = U_0^{\dagger} (\tau';\tau) V\, U_0 (\tau';\tau)$\,. The sum over $\sum_i$ is meant to schematically represent the tracing out of all the environment degrees of freedom (and would, more explicitly, correspond to integrating out of sub-Hubble momentum modes). $L_1$ and $L_2$ are the so-called Lindblad operators. The vectors $\ket{\e_0}$ denote the initial environment state, which we take to coincide with the Bunch-Davies vacuum $\ket{0}$. Likewise, $\ket{\s_0}$ stands for the initial state of the system modes, which are also Bunch-Davies states for modes that start inside the horizon but that eventually cross it to form part of the system (at a given time $\tau$). On the other hand, $\ket{\e_i}$ and $\ket{\s(\tau)}$ denote the environment and system states at time $\tau$, respectively.   

There are a couple of important things we should point out at this stage. Most crucially, note that the form of our equation automatically highlights the main assumption underlying our work -- \textit{Markovian approximation}. In general, a master equation for a reduced density matrix should take the following form \cite{Lindblad:1975ef}:
\begin{eqnarray}
\partial_\tau \rho_\s (\tau) = -i \left[H_{\rm eff}(\tau) , \rho_\s(\tau)\right] +  \sum_k \gamma_k(\tau) \left(L_k \rho_\s(\tau) L_k^\dagger - \frac{1}{2} \left\{L_k^\dagger L_k, \rho_\s(\tau) \right\}\right)\,.
\end{eqnarray}
The above expression is very similar to our master equation Eqn. \eqref{meq0} except that now we have allowed for arbitrary dissipation coefficients $\gamma(t)$, which represent the exchange of energy between the system and environment and is the quintessential marker for the non-unitary evolution of the reduced density matrix. What is crucial here is that under the Markovian approximation, one finds numerical values for $\gamma$ implying that they are constant and $\gamma >0$. This is indeed the case for us and shows how the Markovian approximation is built into our formalism. In fact, we are considering the simplest case where although we are allowing for non-unitary dynamics of the super-Hubble modes (as is typical for open EFTs), we require time-independence of the dissipation coefficients \cite{Prudhoe:2022pte}. Even for a time-local, Markovian system, one might allow for the $\gamma$'s to be time-dependent, and this would imply $\gamma(\tau)>0$ at all times. What we have shown in Appendix \ref{ap:DK} is that our assumption of the Markovian evolution is, at least, self-consistent. There, we have evaluated the dissipation kernel corresponding to our system and shown that it is indeed sharply-peaked, thereby justifying our time-local approximation \cite{BreuerHeinz, boyanovsky2017heisenberg}. Of course, the more interesting and general case would be to consider a time nonlocal master equation but more on this will be discussed in Section \ref{sec:CN}.

Another conceptual point we wish to clarify is the choice of the basis states $\ket{\e_i}$, over which we trace out at some time $\tau$. As has been emphasized in \cite{Gong:2019yyz}, these are not the same as the state one gets from evolving the initial environment state under the free evolution operator (corresponding to the sub-Hubble modes), \textit{i.e.}
\begin{eqnarray}
	\ket{\e_i} \neq  U_{0, \e} (\tau;\tau_0)\, \ket{\e_0}\,.
\end{eqnarray}
Rather, at a given time $\tau$, we identify a complete set of basis states $\ket{\e_i}$, over which we take the trace of the environment modes. Therefore, these are not time-dependent (unlike the right hand side of the above equation). On the other hand, such a choice of a basis for the sub-Hubble modes at a given time $\tau$ indicates that there will be considerable overlap between $\ket{\e_0}$ and $\ket{\e_i}$. The subtleties of this choice for tracing out the environment dofs have been described in detail in \cite{Gong:2019yyz, Gong:2020gdb}. 

\subsection{Perturbative solution}
In order to derive the perturbative solution of the reduced density matrix, we refer to Eqn. (\ref{sol2ndo}) in Appendix \ref{Ame}, and then trace over the environment degrees of freedom, in the same fashion as in Eqn. (\ref{tr1}). Since Eqn.(\ref{sol2ndo}) is given in the interaction picture, we must also multiply each side of it by the free theory unitary operator and its conjugate, as in $\rho(\tau) = U_0 (\tau;\tau_0) \rho_I (\tau) U_0^{\dagger}(\tau;\tau_0)$.

First, let us work at the zeroth-order approximation in some detail, which will help us learn some of the technicalities needed when we go to second order. Using the fact that, at this order $\rho_I(\tau) \approx \rho_I( \tau_0)$, we have
\begin{align}
    \rho_{\rm red} (\tau)  = \sum_i \langle \e_i | \rho_I(\tau_0) | \e_i \rangle & = \sum_i \langle \e_i | U_0 (\tau;\tau_0) \rho (\tau_0) U_0^{\dagger} (\tau;\tau_0) | \e_i \rangle \nonumber \\
    & = \sum_i \langle \e_i | U_{0,\e} | \e_0 \rangle U_{0,\s} | \s_0 \rangle  \langle \s_0 | U_{0,\s}^{\dagger} \langle \e_0 | U_{0,\e}^{\dagger} | \e_i \rangle \nonumber \\
    & = \sum_i \langle \e_i | U_{0,\e} | \e_0 \rangle | \s (\tau) \rangle  \langle \s (\tau) | \langle \e_0 | U_{0,\e}^{\dagger} | \e_i \rangle \nonumber \\
    & = | \s(\tau) \rangle \langle \s(\tau)| \sum_i \langle \e_0 | U_{0,\e}^{\dagger} | \e_i \rangle \langle \e_i | U_{0,\e} | \e_0 \rangle \nonumber \\
    & = \rho_{\s} (\tau) \langle \e_0 | \e_0 \rangle = \rho_\s (\tau)\;,
\end{align}
where we have used $U_0 (\tau;\tau_0) = U_\e (\tau;\tau_0)\, U_\s (\tau;\tau_0)$, and the fact that the states $|\e_i\rangle$ form a complete basis. Further, notice that we have recovered the equation $\rho_{\rm red}^{(0)} (\tau) = \rho_\s (\tau)$, which was used in the previous section to find the master equation. 

Before going to the second-order solution, let us warn the reader we will relax our notation a bit from hereon. As mentioned before, any given operator can be expanded in terms of its sub-horizon and super-horizon Fourier modes as $\hat{A} = \sum_i \hat{A}_{\e}^i \otimes \hat{A}_{\s}^i$. Henceforth, we shall omit the sum and represent every operator as $\hat{A} = \hat{A}_{\e} \otimes \hat{A}_{\s}$. Finally, we introduce the functions $G_I (\tau) = \int d\tau' V_I (\tau')$, where the integrands are either $h_0 (\tau)$ or $h_1(\tau)$, depending on the combination of creation and annihilation operators (as shown in Eqn. \eqref{HI}) in action. With these considerations, schematically the correction to the reduced density matrix at second order is given by
\begin{align}
    \rho_{\rm red}(\tau) \sim - \int_{\tau_0}^{\tau} d\tau' \bigg\{& \left\langle \e_0 \left|  V_{I,\e} (\tau') G_{I,\e} (\tau') \right| \e_0 \right\rangle \left[ U_{0,\s} (\tau;\tau_0) V_{I,\s}(\tau')G_{I,\s}(\tau')U_{0,\s}^{\dagger} (\tau;\tau_0) \rho_\s (\tau) \right. \nonumber \\
    & - \left. U_{0,\s} (\tau;\tau_0) G_{I,\s}(\tau') U_{0,\s}^{\dagger} (\tau;\tau_0) \rho_\s (\tau) U_{0,\s} (\tau;\tau_0) V_{I,\s}(\tau') U_{0,\s}^{\dagger} (\tau;\tau_0) \right] \nonumber \\
    + & \left\langle \e_0 \left| G_{I,\e} (\tau') V_{I,\e}(\tau') \right| \e_0 \right \rangle \left[\rho_\s (\tau) U_{0,\s} (\tau;\tau_0) G_{I,\s}(\tau') V_{I,\s}(\tau') U_{0,\s}^{\dagger} (\tau;\tau_0) \right. \nonumber \\
    & - \left. U_{0,\s}^{\dagger} (\tau;\tau_0) V_{I,\s}(\tau') U_{0,\s}^{\dagger} (\tau;\tau_0) \rho_\s (\tau) U_{0,\s} (\tau;\tau_0) G_{I,\s}(\tau') U_{0,\s}^{\dagger} (\tau;\tau_0) \right] \bigg\}\;.
\end{align}

\section{Power Spectrum}\label{sec:PS}

In this section we will compute the corrections to the tensor power spectrum due to the evolution of the initial pure state into a mixed one, signaling the generation of entanglement entropy \cite{Brahma:2020zpk}. 

\subsection{Linear power spectrum}
As before, we start with the zeroth-order approximation, showing that the present formalism is obviously consistent with the expected outcome. 

First, the tensor power spectrum is defined to be
\begin{equation}
    \Delta^2_t = \frac{q^3}{2\pi^2} \left\langle \hat{h}_{ij}(\vt{q}) \hat{h}_{ji}(-\vt{q}) \right\rangle\,,
\end{equation}
where the two-point function at zeroth order is computed as follows
\begin{align}
   \left\langle \hat{h}_{ij}(\vt{q}) \hat{h}_{ji}(-\vt{q}) \right\rangle & = \Tr\left[ \hat{h}_{ij}(\vt{q}) \hat{h}_{ji}(-\vt{q}) \rho_\s(\tau)\right] \nonumber \\
   & = \Tr\left[ \hat{h}_{ij}(\vt{q}) \hat{h}_{ji}(-\vt{q}) \ket{\s(\tau)}\bra{\s(\tau)}    \right] \nonumber \\
   & = \bra{\s(\tau)} \hat{h}_{ij}(\vt{q}) \hat{h}_{ji}(-\vt{q}) \ket{\s(\tau)} \nonumber \\
   & = \bra{\s_0} U_{0,\s}^{\dagger} (\tau;\tau_0) \hat{h}_{ij}(\vt{q}) \hat{h}_{ji}(-\vt{q}) U_{0,\s} (\tau;\tau_0) \ket{\s_0} \nonumber\\
   & = \bra{\s_0} \hat{h}_{ij}(\tau_,\vt{q}) \hat{h}_{ji}(\tau,-\vt{q}) \ket{\s_0} \nonumber \\
   & = \frac{2}{a^2 \Mp^2} \left[ 4 \left| v_{q} (\tau) \right|^2 \right] = \frac{4}{a^2\Mp^2} \frac{1}{q} \left[1 + \frac{1}{(q\tau)^2}\right]\,.
\end{align}
In this way, obtain
\begin{equation}
    \Delta^2_t = \frac{2}{\pi^2} \frac{H^2}{\Mp^2} (q\tau)^2 \left[1 + \frac{1}{(q\tau)^2}\right] \approx \frac{2}{\pi^2} \frac{H^2}{\Mp^2}\;,
\end{equation}
as expected at late times $\tau \rightarrow 0$. We show this simple derivation in detail to explain to the reader what we operationally mean by taking the trace over the environment dofs, as it will be done later on for the first order quantum correction in a more complicated context. 

\subsection{Corrections to the power spectrum: Seeting up the problem}\label{PCPP}
Firstly, for the sake of brevity we abbreviate the product of tensor modes as
\begin{equation}
    \hat{{\cal O}}_{\vt q} := \hat{h}_{ij}(\vt{q}) \hat{h}_{ji}(-\vt{q})\;,
\end{equation}
which takes the following form in the interaction picture:
\begin{align*}
    {\hat {\cal O}}_{\vt q} (\tau) = \frac{2}{a^2 \Mp^2} \sum_{\alpha} & \Big\{ e_{ij}^{\alpha} ({\vt q}) e_{ji}^{\alpha} (- {\vt q}) \left[ v_q (\tau) \right]^2 \aop{a}{q}^{\alpha} \aop{a}{- \vt q}^{\alpha} + e_{ij}^{\alpha} ({\vt q}) e_{ji}^{\alpha} ({\vt q}) \left| v_{q} (\tau) \right|^2 \aop{a}{q}^{\alpha} \aop{a}{q}^{\alpha \dagger} \nonumber \\
    & + e_{ij}^{\alpha} (- {\vt q}) e_{ji}^{\alpha} (- {\vt q}) \left| v_q (\tau) \right|^2 \aop{a}{-q}^{\alpha \dagger} \aop{a}{-q}^{\alpha} + e_{ij}^{\alpha} (- {\vt q}) e_{ji}^{\alpha} ({\vt q}) \left[ v_q^* (\tau) \right]^2 \aop{a}{-q}^{\alpha \dagger} \aop{a}{q}^{\alpha \dagger}            \Big\}\;.
\end{align*}
Even though we are working in the Schr\"odinger picture, the resulting combination of unitary operators will take $\hat{{\cal O}}_{\vt q}$ into $\hat{{\cal O}}_{\vt q} (\tau)$. Furthermore, notice that the creation and annihilation operators act on the initial (Bunch-Davies) vacuum. Considering this, the correction to the tensor spectrum is of the form
\begin{align}
    \Tr \left[ \hat{\cal O}_{\vt q} \rho_{\rm red}^{(2)} (\tau) \right] =\ & (-i)^2 \int_{\tau_0}^{\tau} d\tau' \bigg\{ \left\langle \e_0 \left|  V_{I,\e} (\tau') G_{I,\e} (\tau') \right| \e_0 \right\rangle \left[ \bra{\s_0} {\hat {\cal O}}_{\vt q} (\tau) V_{I, \s} (\tau') G_{I, \s}(\tau') \ket{\s_0} \right. \nonumber \\
    & \left. - \bra{\s_0} V_{I,\s}(\tau') {\hat {\cal O}}_{\vt q} (\tau) G_{I,\s} (\tau') \ket{\s_0} \right] + \left\langle \e_0 \left| G_{I,\e} (\tau') V_{I,\e}(\tau') \right| \e_0 \right \rangle \times \nonumber \\
    & \left[ \bra{\s_0} G_{I,\s}(\tau') V_{I,\s}(\tau') {\hat {\cal O}}_{\vt q} (\tau) \ket{\s_0} - \bra{\s_0} G_{I,\s}(\tau') {\hat {\cal O}}_{\vt q} (\tau) V_{I,\s}(\tau') \ket{\s_0} \right] \bigg\} \nonumber \\
    =\ & (-i)^2 \int_{\tau_0}^{\tau} d\tau' \bigg\{ \left\langle \e_0 \left|  V_{I,\e} (\tau') G_{I,\e} (\tau') \right| \e_0 \right\rangle \left[ \bra{\s_0} {\hat {\cal O}}_{\vt q} (\tau) V_{I, \s} (\tau') G_{I, \s}(\tau') \ket{\s_0} \right. \nonumber \\
    & \left. - \bra{\s_0} V_{I,\s}(\tau') {\hat {\cal O}}_{\vt q} (\tau) G_{I,\s} (\tau') \ket{\s_0} \right] + {\rm c.c.} \bigg\}\;. \label{trr2}
\end{align}

In principle, we now must deal with every possible combination of system and environment modes in both $V_I$ and $G_I$. Nevertheless, we must have at least one system mode, otherwise we are missing the effect of the environment on the system. Then, we would have to work with any possible combination of $(\e\e\s)$ and $(\e\s\s)$ modes. Notice that due to momentum conservation (imposed by the delta functions), the sum of the three momenta must be 0. This constrains the contribution from the $(\e\s\s)$ combination (which can be thought of being of the `folded' shape), and is thus subdominant in comparison to the `squeezed limit' contribution $(\e\e\s)$. 

\subsubsection{Inner products}

The resulting scalar products in Eqn. \eqref{trr2} determine which combinations of creation and annihilation operators contribute to the corrections to the spectrum. Both $V_I$ and $G_I$ carry every possible combination of the product of three of such operators ($\hat{c}$ and $\hat{c}^{\dagger}$), as shown in Eqn. \eqref{HI} for $V_I$ and its time integration for $G_I$. However, since we have to deal with scalar products on different Hilbert spaces (${\cal H}_\e$ for the environment and ${\cal H}_\s$ for the system), the creation and annihilation operators must act on each space depending on the magnitude of the momentum associated with them. For notational convenience, when $\hat{c}$ ($\hat{c}^{\dagger}$) acts on ${\cal H}_\e$ we shall denote it by $\hat{b}$ ($\hat{b}^{\dagger}$), and by $\hat{a}$ ($\hat{a}^{\dagger}$) when it acts on ${\cal H}_\s$. In this way, the one–particle states in the system and environment at the time $\tau$ are respectively given by $\hat{a}^{\alpha \dagger}_{\vt k}\ket{\s_0} = \ket{1_{\vt k}^{\alpha}}$ if $k < aH$; and $\hat{b}^{\alpha \dagger}_{\vt k} \ket{\e_0} = \ket{1_{\vt k}^{\alpha}}$ if $k > aH$.

Looking at the scalar product on ${\cal H}_{\e}$ in Eqn. \eqref{trr2}, we can see that the combinations that contribute to the product must be $V_{I}(\tau') \sim \aop{b}{} \aop{b}{} \aop{a}{}$ or $\sim \aop{b}{} \aop{b}{} \aop{a}{}^{\dagger}$, and $G_{I} \sim \aop{b}{}^{\dagger} \aop{b}{}^{\dagger} \aop{a}{}$ or $\sim \aop{b}{}^{\dagger} \aop{b}{}^{\dagger} \aop{a}{}^{\dagger}$.\footnote{Notice that combinations of the form $V_I \sim \hat{b} \hat{b}^{\dagger} \hat{a}$ are also possible. However, due to momentum conservation, the momentum associated with $\hat{a}$ (or $\hat{a}^{\dagger}$) can only be zero.}
From the expression for $\hat{\cal O}_{\vt q}(\tau)$ and from the trace, we can see that the kind of terms we need to compute are of the form
\begin{align}
   & \int_{\Delta k}\int_{\Delta p}\Big\{ e_{ij}^{\alpha} ({\vt q}) e_{ji}^{\alpha} (- {\vt q}) \left[ v_q (\tau) \right]^2 \bra{1_{\vt q}^{\alpha}, 1_{- \vt q}^{\alpha}} \aop{a}{-k_1}^{\alpha_1 \dagger} \aop{a}{-p_1}^{\beta_1 \dagger} \ket{\s_0} 
   h_1 (\tau', -{\vt k_1}, {\vt k_2}, {\vt k_3}) h_0^* (\tau'', -{\vt p_1}, -{\vt p_2}, -{\vt p_3}) \nonumber \\
   +\ & e_{ij}^{\alpha} ({\vt q}) e_{ji}^{\alpha} ({\vt q}) \left| v_{q} (\tau) \right|^2 \bra{\s_0} \aop{a}{k_1}^{\alpha_1} \aop{a}{-p_1}^{\beta_1 \dagger}\ket{\s_0} h_0 (\tau', {\vt k_1}, {\vt k_2}, {\vt k_3}) h_0^* (\tau'', -{\vt p_1}, -{\vt p_2}, -{\vt p_3})  \nonumber \\
&  -\bra{1_{\vt k_1}^{\alpha_1}} {\hat {\cal O}}_{\vt q} (\tau) \ket{1_{-\vt p_1}^{\beta_1}} h_0(\tau', {\vt k_1}, {\vt k_2}, {\vt k_3}) h_0^* (\tau'', -{\vt p_1}, -{\vt p_2}, -{\vt p_3}) \Big\} \left\langle 1_{\vt k_2}^{\alpha_2}, 1_{\vt k_3}^{\alpha_3} \right| \left. 1_{-\vt p_2}^{\beta_2}, 1_{-\vt p_3}^{\beta_3} \right\rangle\,,
\end{align}
where we have fixed ${\vt k_1}$ and ${\vt p_1}$ for the system modes and the remaining momenta for the environment modes. In the end, we will consider every case by multiplying the result by an appropriate multiplicity factor. 

Next, let us tackle each term above separately, where the first one is simply given by
\begin{align*}
    & e_{ij}^{\alpha} ({\vt q}) e_{ji}^{\alpha} (- {\vt q}) \left[ v_q (\tau) \right]^2 \left[ h_1 (\tau', -{\vt q}, {\vt k_2}, - {\vt q} - {\vt k_2}) h_0^* (\tau'', \vt{q}, {\vt k_2}, -{\vt q} - {\vt k_2}) \right. \nonumber \\
    & + \left. h_1 (\tau', {\vt q}, {\vt k_2}, {\vt q} - {\vt k_2}) h_0^* (\tau'', -\vt{q}, {\vt k_2}, {\vt q} - {\vt k_2}) \right] \delta_{\alpha, \alpha_1}\delta_{\alpha,\beta_1}\,.
\end{align*}
The second term does not present a `contraction' between $\vt{q}$ and $\vt{k_1}$, and thus we've got
\begin{align} \label{eq2nd}
    \sum_{\vt k_1} e_{ij}^{\alpha} ({\vt q}) e_{ji}^{\alpha} ({\vt q}) \left| v_{q} (\tau) \right|^2  h_0(\tau', {\vt k_1}, {\vt k_2}, -{\vt k_1} - {\vt k_2}) h_0^* (\tau'', {\vt k_1}, {\vt k_2}, -{\vt k_1} - {\vt k_2}) \delta_{\alpha_1, \beta_1}\,.
\end{align}
The final term is the most complicated one, so we break it in parts as follows
\begin{align*}
     &\sum_{\vt k_1}\bra{1_{\vt k_1}^{\alpha_1}} {\hat {\cal O}}_{\vt q} (\tau) \ket{1_{\vt k_1}^{\beta_1}} h_0(\tau', {\vt k_1}, {\vt k_2}, {\vt k_3})  h_0^* (\tau'', {\vt k_1}, {\vt k_2}, {\vt k_3}) = \nonumber \\
     & \bra{1_{\vt q}^{\alpha_1}} {\hat {\cal O}}_{\vt q} (\tau) \ket{1_{\vt q}^{\beta_1}} h_0(\tau', {\vt q}, {\vt k_2}, {\vt k_3}) h_0^* (\tau'', {\vt q}, {\vt k_2}, {\vt k_3}) \nonumber \\
    & + \bra{1_{-\vt q}^{\alpha_1}} {\hat {\cal O}}_{\vt q} (\tau) \ket{1_{-\vt q}^{\beta_1}} h_0(\tau', -{\vt q}, {\vt k_2}, {\vt k_3}) h_0^* (\tau'', -{\vt q}, {\vt k_2}, {\vt k_3}) \nonumber \\
    &+ \sum_{{\vt k_1} \neq \pm {\vt q}}\bra{1_{\vt k_1}^{\alpha_1}} {\hat {\cal O}}_{\vt q} (\tau) \ket{1_{\vt k_1}^{\beta_1}} h_0(\tau', {\vt k_1}, {\vt k_2}, {\vt k_3}) h_0^* (\tau'', {\vt k_1}, {\vt k_2}, {\vt k_3}) \nonumber \\
    =\ & 2 e_{ij}^{\alpha} ({\vt q}) e_{ji}^{\alpha} ({\vt q}) \left| v_{q} (\tau) \right|^2 h_0(\tau', {\vt q}, {\vt k_2}, -{\vt q} - {\vt k_2}) h_0^* (\tau'', {\vt q}, {\vt k_2}, -{\vt q} - {\vt k_2}) \delta_{\alpha, \alpha_1}\delta_{\alpha, \beta_1} \nonumber \\
    &+ e_{ij}^{\alpha} ({\vt q}) e_{ji}^{\alpha} ({\vt q}) \left| v_{q} (\tau) \right|^2 h_0(\tau', {\vt q}, {\vt k_2}, -{\vt q} - {\vt k_2}) h_0^* (\tau'', {\vt q}, {\vt k_2}, -{\vt q} - {\vt k_2}) \delta_{\alpha_1, \beta_1} (1-\delta_{\alpha, \alpha_1}) \nonumber \\
    &+ e_{ij}^{\alpha} ({\vt q}) e_{ji}^{\alpha} ({\vt q}) \left| v_{q} (\tau) \right|^2 h_0(\tau', -{\vt q}, {\vt k_2}, {\vt q} - {\vt k_2}) h_0^* (\tau'', -{\vt q}, {\vt k_2}, {\vt q} - {\vt k_2}) \delta_{\alpha_1, \beta_1} \nonumber \\
    &+ e_{ij}^{\alpha} (-{\vt q}) e_{ji}^{\alpha} (-{\vt q}) \left| v_{q} (\tau) \right|^2 h_0(\tau', -{\vt q}, {\vt k_2}, {\vt q} - {\vt k_2}) h_0^* (\tau'', -{\vt q}, {\vt k_2}, {\vt q} - {\vt k_2}) \delta_{\alpha,\alpha_1} \delta_{\alpha, \beta_1} \nonumber \\
    &+ e_{ij}^{\alpha} ({\vt q}) e_{ji}^{\alpha} ({\vt q}) \left| v_{q} (\tau) \right|^2 \sum_{{\vt k_1} \neq \pm {\vt q}} h_0(\tau', {\vt k_1}, {\vt k_2}, -{\vt k_1} - {\vt k_2}) h_0^* (\tau'', {\vt k_1}, {\vt k_2}, -{\vt k_1} - {\vt k_2}) \delta_{\alpha_1, \beta_1}\,.
\end{align*}
Notice that, from the last equality, we can combine one factor from the first line, together with the second, third and last lines to cancel out with the expression from the second term above, Eqn. \eqref{eq2nd}, and thus we end up with
\begin{align}\label{corx}
    & \Tr \left[ \hat{\cal O}_{\vt q} \rho_{\rm red}^{(2)} (\tau) \right] = - \frac{2}{a^2 \Mp^2} \times 18 \int_{-1/q}^{\tau} d\tau' \int_{\e} \frac{d^3 k_2}{(2\pi)^3}\int_{-1/q}^{\tau'} d\tau''  \bigg\{ e_{ij}^{\alpha} ({\vt q}) e_{ji}^{\alpha} (- {\vt q}) \left( v_q (\tau) \right)^2 \nonumber \\   
    & \bigg[ h_1 (\tau', -{\vt q}, {\vt k_2}, - {\vt q} - {\vt k_2}) h_0^* (\tau'', \vt{q}, {\vt k_2}, -{\vt q} - {\vt k_2}) 
     + h_1 (\tau', {\vt q}, {\vt k_2}, {\vt q} - {\vt k_2}) h_0^* (\tau'', -\vt{q}, {\vt k_2}, {\vt q} - {\vt k_2}) \bigg] \delta_{\alpha, \alpha_1}\delta_{\alpha,\beta_1} \nonumber \\
    & - \bigg[ e_{ij}^{\alpha} ({\vt q}) e_{ji}^{\alpha} ({\vt q}) h_0 (\tau', {\vt q}, {\vt k_2}, -{\vt q} - {\vt k_2}) h_0^* (\tau'', {\vt q}, {\vt k_2}, -{\vt q} - {\vt k_2}) \nonumber \\
    & + e_{ij}^{\alpha} (- \vt{q}) e_{ji}^{\alpha} (-\vt{q}) h_0(\tau', -{\vt q}, {\vt k_2}, {\vt q} - {\vt k_2}) h_0^* (\tau'', -{\vt q}, {\vt k_2}, {\vt q} - {\vt k_2}) \bigg] \left| v_q (\tau) \right|^2 \delta_{\alpha, \alpha_1}\delta_{\alpha, \beta_1} + {\rm c.c.} \bigg\} \;,
\end{align}
where 18 is the multiplicity factor (3 ways of choosing the system mode in $h_0$, 3 factors from $h_1$, and 2 ways of contracting the inner product over environment modes).

\subsubsection{Contribution of different polarizations to the integral}
Each product above contains complicated sums over polarizations, for which we show the results below. Notice that due to the state (scalar) products, the first momenta on the left has the same polarization as the first on the right, and so on. In consequence, the sums of interest are, upon integration over $\phi$,
\begin{align}
     \sum_{\{\alpha_i\}} e_{ij}^{\alpha_1} (\vt{q}) e_{ji}^{\alpha_1} (-\vt{q}) e_{i_1 j_1}^{\alpha_1} (-\vt{q}) e_{j_1 k_1}^{\alpha_2} (\vt{k_2}) e_{k_1 i_1}^{\alpha_3} (-\vt{k_2}-\vt{q}) e_{i_2 j_2}^{\alpha_1} (\vt{q}) e_{j_2 k_2}^{\alpha_2} (\vt{k_2}) e_{k_2 i_2}^{\alpha_3} (-\vt{k_2}-\vt{q}) = \nonumber \\
    2\pi\frac{\left[4q^4 + 4k_2^4 + 11 q^2 k_2^2 + 8 q k_2 (q^2 + k_2^2) \cos \theta + q^2 k_2^2 \cos (2\theta) \right] \sin^4 \theta}{(q^2 + k_2^2 + 2q k_2 \cos \theta)^2} := f_1 (\theta)\,, \label{eqf1}
\end{align}
\begin{align}
     \sum_{\{\alpha_i\}} e_{ij}^{\alpha_1} (\vt{q}) e_{ji}^{\alpha_1} (-\vt{q}) e_{i_1 j_1}^{\alpha_1} (\vt{q}) e_{j_1 k_1}^{\alpha_2} (\vt{k_2}) e_{k_1 i_1}^{\alpha_3} (-\vt{k_2}+\vt{q}) e_{i_2 j_2}^{\alpha_1} (-\vt{q}) e_{j_2 k_2}^{\alpha_2} (\vt{k_2}) e_{k_2 i_2}^{\alpha_3} (-\vt{k_2}+\vt{q}) = \nonumber \\
    2\pi\frac{\left[4q^4 + 4k_2^4 + 11 q^2 k_2^2 - 8 q k_2 (q^2 + k_2^2) \cos \theta + q^2 k_2^2 \cos (2\theta) \right] \sin^4 \theta}{(q^2 + k_2^2 - 2q k_2 \cos \theta)^2}  := f_2 (\theta)\,,
\end{align}
\begin{align}
     \sum_{\{\alpha_i\}} e_{ij}^{\alpha_1} (\vt{q}) e_{ji}^{\alpha_1} (\vt{q}) e_{i_1 j_1}^{\alpha_1} (\vt{q}) e_{j_1 k_1}^{\alpha_2} (\vt{k_2}) e_{k_1 i_1}^{\alpha_3} (-\vt{k_2}-\vt{q}) e_{i_2 j_2}^{\alpha_1} (\vt{q}) e_{j_2 k_2}^{\alpha_2} (\vt{k_2}) e_{k_2 i_2}^{\alpha_3} (-\vt{k_2}-\vt{q}) = f_1 (\theta)\;,
\end{align}
\begin{align}
     \sum_{\{\alpha_i\}} e_{ij}^{\alpha_1} (-\vt{q}) e_{ji}^{\alpha_1} (-\vt{q}) e_{i_1 j_1}^{\alpha_1} (-\vt{q}) e_{j_1 k_1}^{\alpha_2} (\vt{k_2}) e_{k_1 i_1}^{\alpha_3} (-\vt{k_2}+\vt{q}) e_{i_2 j_2}^{\alpha_1} (-\vt{q}) e_{j_2 k_2}^{\alpha_2} (\vt{k_2}) e_{k_2 i_2}^{\alpha_3} (-\vt{k_2}+\vt{q}) = f_2 (\theta)\;, \label{eqf2}
\end{align}
where without loss of generality we have considered 
\begin{equation}
    \vt{q} = q(0,0,1) = \pm \vt{k_1}\,, \quad \vt{k_2} = k_2(\sin \theta \cos \phi, \sin \theta \sin \phi, \cos \theta)\,,
\end{equation}
and due to momentum conservation $\vt{k_3} = -(\vt{k_1} + \vt{k_2})$. Notice that the momentum we use can be $\pm \vt{k_3}$ depending on whether we used a creation or an annihilation operator.  
Then, we need to solve integrals of the type (say, for the third term on Eqn.\eqref{corx}),
\begin{align}
    \frac{q^3}{2\pi^2}\frac{36}{a^2 \Mp^2}\frac{32H^2}{\Mp^2} |v_q (\tau)|^2  \int_{-1/q}^{\tau} \frac{d\tau'}{\tau'} \int_{-1/\tau'}^{\infty} \frac{dk_2}{(2\pi)^3}k_2^2 &\int_{-1}^{1} d(\cos \theta) f_1(\cos \theta) v_q(\tau')v_{k_2}(\tau') v_{k_3}(\tau')\nonumber \\
    \times &\int_{-1/q}^{\tau'} \frac{d\tau''}{\tau''}   [v_q(\tau'')]^* [v_{k_2}(\tau'')]^* [v_{k_3}(\tau'')]^*\;. \label{3rdl}
\end{align}

\subsubsection{Integration over time and momenta}

The correction to the spectrum follows from the rather complicated integral shown above and other terms which are similarly very involved. In principle, one can try and play with the order of integrations to facilitate the task ahead, although special care has to be put into the integration limits to be consistent with which modes belong to the environment and at which times. So, in principle, we could perform first the integration over the environment mode $k_2$, which allows to define a dissipation kernel in an approach similar to \cite{Boyanovsky:2015tba, Brahma:2021mng}. We try and follow this path in Appendix \ref{ap:DK}, which leads to a dead-end due to the complicated form of the remaining integrals (however, the kernel can provide useful information about the physics of the system and, in particular, show its time locality). To proceed further in that vein would require numerical solutions. Here, we shall follow a more fruitful path, where we start with the integration over $\tau''$ and be able to evaluate our result analytically.

In order to deal with the annoying pre-factors of the mode functions, let us define $\beta_k (\tau) := \sqrt{2k}\ v_k (\tau)$, such that the integral takes the form
\begin{align*}
    \frac{9}{2\pi^5} \frac{H^4}{\Mp^4} q\tau^2 \left(1 + \frac{1}{(q\tau)^2} \right)\int_{-1/q}^{\tau} \frac{d\tau'}{\tau'} \int_{-1/\tau'}^{\infty} dk_2 \int_{-1}^{1} d(\cos \theta) \frac{k_2}{k_3} f_1(\cos \theta) \beta_q(\tau')\beta_{k_2}(\tau') \beta_{k_3}(\tau') \Xi (\tau', q, k_2,\cos \theta)\;,
\end{align*}
where $\Xi$ is the function resulting from the integration over $\tau''$, and is given by
\begin{align}
    \Xi = & \frac{1}{3 q k_2 k_3} \Bigg\{ iq e^{-i (q + k_2 + k_3)/q} \Big[ 2 k_2^2 - (1-i) q k_2 + (3+i)q^2 + 2 k_2 q y - k_2 k_3 - (1-i) q k_3 \Big]  \nonumber \\
    & + \frac{e^{i (q + k_2 + k_3)\tau'}}{\tau'^3} \Big[ i + 2i k_2^2 \tau'^2 + \big[q + k_3 - iq \tau' (k_3 - 2q) \big]\tau' + k_2 \tau' \big[ 1 - i (q (1-2y) + k_3)\tau' \big] \Big] \nonumber \\
    & - \Big[ k_2^3 + k_2^2 k_3 + 2 q k_2 k_3 y + q^2 (q + k_3) \Big] \Big[ \text{Ei} \left( - i (q + k_2 + k_3)/q \right) - \text{Ei} \left( i(q + k_2 + k_3) \tau' \right) \Big] \Bigg\}\;,
\end{align}
with $y := \cos \theta$. Next, we introduce the dimensionless variables $\omega' := -q \tau'$ and $\kappa_i := -k_i \tau'$, such that the integral Eqn. \eqref{3rdl} becomes
\begin{align}
   \Delta^2_t \sim \frac{9}{2\pi^5} \frac{H^4}{\Mp^4} (1 + \omega^2)\int_{1}^{\omega} \frac{d\omega'}{(\omega')^2} \int_{1}^{\infty} d\kappa_2  \int_{-1}^{1} dy\ \frac{\kappa_2}{\kappa_3} f_1(y) \beta(\omega')\beta(\kappa_2) \beta(\kappa_3) \Xi (\omega', \kappa_2, y)\;.
\end{align}
Then, we Taylor expand the entire integrand around $\kappa_2 \rightarrow \infty$, which renders
\begin{align}
   \sim \frac{2\pi (y^2-1)^2}{\kappa_2^2\omega'^4}&\Big\{ 2 i \kappa_2 (1+\omega'^2) - 7 i y \omega' (1 + \omega'^2) - 2 - \omega'^2 - i \omega'^3 + \exp \left[ \frac{i (\omega'-1)}{\omega'} \left(2\kappa_2 + (1+y)\omega' \right) \right] \nonumber\\
  & \times \omega'^2 (i + \omega') \left[4(1-i) - 2(1+i)\kappa_2 + \left( (-2 + 3i) + 7(1+i)y \right)\omega' \right]   \Big\}\;.
\end{align}
This is a good approximation for large values of $\kappa_2$, and a reasonable one for $\kappa_2 \sim 1$. Then, we can integrate over $y$, $\kappa_2$ and $\omega'$ (in that order), to find
\begin{align}\label{prefinal}
  \Delta^2_t \sim \frac{9}{2\pi^5} \frac{H^4}{\Mp^4} \omega^2 & \Bigg\{ -\left(1+\frac{1}{\omega^2}\right) \frac{64 i \pi}{45} \left(\frac{1}{\omega^{3}} + \frac{3}{\omega} - 4\right) \ln \frac{\Lambda}{H} + \left(1+\frac{1}{\omega^2}\right) \frac{64 i \pi}{45} \left(\frac{1}{\omega^{3}} + \frac{3}{\omega} - 4\right) \ln \frac{\Lambda}{H} \nonumber 
  \\ & + \frac{128\pi}{45\omega^2} \left( \frac{1}{\omega^3} - \frac{1}{2 \omega^3} [3 - 90\pi(1+5 \cos 2 + \sin 2)] + \frac{1}{\omega} - 1 \right) \Bigg\}\;,
\end{align}
where we have also included the complex conjugate term. 

The terms on the first line come from the upper limit of the $\kappa_2$ momentum integration, where $a \Lambda$ represents a comoving cutoff for $k_2$ which translates into $\Lambda/H$ for $\kappa_2 \,(= -k_2/(a(\tau') H)$, whereas the terms on the second line come from $\kappa_2 \rightarrow 1$ (for $k_2$, this lower limit would have been $aH$). Notice that the UV-divergences (going as $\ln \Lambda$) have been canceled out due to adding in the complex conjugate (they appear as a purely imaginary contribution). However, that is not the case for the divergences for all the terms, for instance, the  logarithmic ones coming from both terms on the second line of Eqn. \eqref{corx}. We have more to say on this later.

\subsubsection{Quantum corrections}
We have now all the ingredients to compute the final result describing the quantum corrections to the tensor power spectrum. Let us state the final result now and then discuss some of the intermediate steps as well as the consequences in the next subsection. 

Following the same prescription as before, we can compute the contribution to the spectrum from each term in Eqn. \eqref{corx}, including their complex conjugates. In doing so, we find that the expressions coming from the limit $\kappa_2  = 1$ cancel out up to order ${\cal O}(\omega)$ (at least). In other words, our final result depends on positive powers of $\omega$ which, in the late time limit $\tau \rightarrow 0$, goes to zero. This is quite remarkable since, looking at Eqn. \eqref{prefinal}, one would have been skeptical that the terms with inverse powers of $\omega$ might have survived and that would have been disastrous. However, they all cancel out neatly in the end.

Then, the only surviving terms are those associated with the UV-divergences, and a rather innocuous IR one as well. We will discuss these terms next, but for now our final result for the leading order correction to the spectrum is given by:
\begin{equation}\label{Final}
    \Delta^2_t \simeq -\frac{256}{5\pi^4} \left(\frac{H}{\Mp}\right)^4 \left\{\big[2 + \cos 2 + \text{Ci}\ 2 - \sin 2 \big] \ln \left(\frac{H}{\mu}\right) + \mathcal{O}(1)\right\}\;.
\end{equation}
where  Ci is the cosine integral function, and we have not explicitly written some $\mathcal{O}(1)$ numerical factors  above (involving the Euler–Mascheroni constant $\gamma_E$ and other such terms) which depend on the renormalization scheme.

\subsection{UV and IR divergences}
Admittedly, we have pulled our final result for the first order quantum correction to the (dimensionless) tensor power spectrum, in Eqn. \eqref{Final}, rather out of the hat. The gruesome details of each of the integrals in Eqn. \eqref{corx} have been shown in Appendix \ref{tpi}. Let us elaborate on some of the intermediate steps now. 

Firstly, note that we have multiplied by a factor of $q^3$ to get the dimensionless power spectrum. A factor of $q^2$ then combines with the factor of $a^2$ from the prefactor to give us a multiplicative factor of $\omega^2 H^2$ while one factor of the leftover $q$ had been used to convert the measure from $d q$ to $d \omega$. After evaluating the integrals, the results can have two types of divergences as usual -- UV and IR ones. To understand them fully, look at our final expressions for the four terms we get from performing these integrals, given in Eqns. \eqref{UV1} -\eqref{Tame2} in Appendix \ref{tpi}. 

For the last two terms  Eqns. \eqref{Tame1} and \eqref{Tame2}, the beautiful cancellations happen as shown in Eqn. \eqref{prefinal} above. The UV divergent terms appear as $i \ln \Lambda$, and therefore, fall out of the final expression due to adding in its complex conjugate.\footnote{These cancellations have already been taken into account in Eqns. \eqref{Tame1} and \eqref{Tame2} by adding in the complex conjugate.} This is not strange at all -- all this suggests is that although we need to put in a cutoff to regulate momentum integrals in the intermediate steps, the physical result is independent of such cutoffs simply due to how the UV divergences appears in them.

On the other hand, there are a plethora of inverse $\omega$ terms remaining in these two expressions which would lead to serious IR divergences when taking the late time limit of $\tau \rightarrow 0$. As mentioned in passing above for Eqn. \eqref{prefinal}, all of these terms cancel and we have a resulting expression which is $\mathcal{O}(\omega^2)$ or higher for these terms. Note that the fact that these terms drop out only in the $q \tau \rightarrow 0$ limit, and not otherwise, does not tell us anything deep at all. Of course, even the linear power spectrum is only scale-invariant in this limit, which is the relevant one for observations. At the risk of being pedantic, we repeat this trivial argument to convince the reader that nothing strange in going on for the quantum correction in this aspect of it. In conclusion, these two terms \eqref{Tame1} and \eqref{Tame2} do not have any UV or IR divergences at all. 

Let us now come to the more interesting terms given in Eqns. \eqref{UV1} and \eqref{UV2}. These terms can be represented as giving a correction of the following generic form (we have omitted a constant prefactor to simplify the expression):
\begin{align}\label{fincon}
    \Delta^2_t \sim  \frac{H^4}{\Mp^4} \big[2 - \gamma_E + \cos 2 + \text{Ci}\ 2 - \sin 2 - \ln (2\omega) \big] \ln \frac{\Lambda}{H}\;,
\end{align}
where we have only written down the terms which survive after several cancellations amongst themselves from Eqns. \eqref{UV1}, \eqref{UV2}, \eqref{Tame1} and \eqref{Tame2} in Appendix - \ref{tpi}. Let us first talk about the UV sensitivity of the above term and deal with the $\ln(2\omega)$ term later. The UV divergence is logarithmic and of the form $\ln\left(\Lambda/H\right)$.\footnote{As mentioned, more severe power-law UV divergences simply never show up in the final expression.} This is what is commonly expected in loop corrections for inflationary perturbations. The common way to deal with them is separate out a divergent piece from the above term and keep the finite contribution, for some `renormalization scale' at which the observations are made \cite{Senatore:2009cf, Senatore:2012ya, Tan:2019czo}. A more rigorous way to achieve the same would be to instead use a dimensional regularization scheme instead of the comoving cut-off used here. However, there is a simple way to formalize the equivalence between the dimensional and cutoff regularizations \cite{Cynolter:2015lea},
\begin{eqnarray}
\frac{1}{\varepsilon} - \gamma_E + \ln (4\pi \mu^2) + 1 = \ln \Lambda^2\;. 
\end{eqnarray}
This equivalence, proven in the context of Gauge theories, shows that diffeomorphism invariance is respected when a comoving cutoff has been introduced in the calculation. The divergent part is isolated $\varepsilon \rightarrow 0$ and can be dispensed with. Up to finite factors that can be absorbed into the divergent part, and up to $\mathcal{O}(1)$ constants depending on the renormalization scheme, the correction to tensor spectrum now reads
\begin{equation}
        \Delta^2_t \sim - \left(\frac{H}{\Mp}\right)^4 \left\{\big[2 + \cos 2 + \text{Ci}\ 2 - \sin 2 \big] \ln \left(\frac{H}{\mu}\right) + \mathcal{O}(1)\right\}\;
\end{equation}
The introduction of the renormalization scale is quite well-known, and has been discussed in great detail in \cite{Senatore:2009cf, Senatore:2012ya, Tan:2019czo, Tanaka:2013caa}. Instead to rehashing those arguments, let us point out that the crucial observation regarding the UV behaviour of our calculation goes as follows. Had we found a term which goes as $\ln(\Lambda/q)$ \cite{Adshead:2009cb, Weinberg:2006ac, Seery:2007we, Sloth:2006az}, that would have been a truly problematic term. As was first pointed out in \cite{Senatore:2009cf}, and has been repeated many times since \cite{Senatore:2012nq, Senatore:2012ya}, these type of terms are simply ruled out by the diffeomorphism symmetry of the problem. However, choosing the correct covariant regularization scheme led us to find a term of the form $\ln (\Lambda/H)$ and this is indeed what is expected from loop corrections. 

However, there is another type of large logarithms appearing in Eqn. \eqref{fincon} coming from the terms Eqns. \eqref{UV1} and \eqref{UV2} in Appendix \ref{tpi}. This is the term proportional to $\ln(|q\tau|)$, which diverges in the IR, when taking the late time limit $\tau \rightarrow 0$. As mentioned, this is \textit{not} a UV divergent term, but rather an IR one.\footnote{The final integral over $\omega$ which gives this term runs from $1$ to $0$, and so there is no UV-limit, \textit{i.e.} $\omega \rightarrow \infty$, to be considered here.} These type of divergences first appeared in \cite{Giddings:2010nc, Byrnes:2010yc, Seery:2010kh} but have been since also shown to be harmless in the sense that they can never affect local observations \cite{Senatore:2012ya, Giddings:2011zd, Gerstenlauer:2011ti}. The essence of the argument is that and the long wavelength can be reabsorbed, through the remaining gauge freedom, by change of coordinates (see \cite{Tanaka:2013caa} for a nice review of these arguments). In \cite{Pimentel:2012tw}, a similar logic was used in the context of scalar perturbations, where the time dependence induced by a cubic interaction (with spatial derivatives) is suppressed through the effects of a quartic interaction necessary to maintain diffeomorphism invariance at late times. Finally, this IR growth appearing in the quantum correction to the tensor power spectrum can be related to the issue of the IR divergence of the Bunch-Davies vacuum for the free graviton mode \cite{Higuchi:2011vw} and is certainly not relevant for any local observable.

In conclusion, what we find is that taking the limit $q \tau \rightarrow 0$ does \textit{not} lead to any divergence at all. This is all the more surprising since, naively, it did look like the perturbative quantum corrections (\textit{e.g.} see Eqn. \eqref{prefinal}) would give large corrections in this limit. However, the fact that the various terms in Eqn. \eqref{Final} nicely cancel out is not an accident at all and, we  believe, is sensitively dependent on our \textit{Markovian approximation}. We will explain this is more detail in the next section.

\section{Conclusions}\label{sec:CN}
\subsection{Interpretation of our result}
The most obvious conclusion to be drawn from our computation is that the leading order quantum corrections to the tensor power spectrum matches exactly with what is expected from loop corrections, when considering purely cubic tensor interactions. Let us elaborate on this statement a bit. Firstly, just from power counting one expects that the one-loop effect will be suppressed by a factor of $\mathcal{O}(H^2/\Mp^2)$. Furthermore, more recent computations of such loop corrections to the tensor power spectrum (not from tensor loops but rather from other forms of matter loops) indicate that we should expect a correction of the form \cite{Tan:2019czo, Frob:2012ui}
\begin{eqnarray}
	\left\langle h_{ij} h_{ij}\right\rangle \sim \frac{1}{k^3}\, \left(\frac{H}{\Mp}\right)^4\,\log\left(\dfrac{H}{\mu}\right)\,,
\end{eqnarray}
where $\mu$ is some renormalization scale. Note that earlier works predicted much larger corrections where the logarithm would contain a term of the form $\log\left(k/\mu\right)$ \cite{Adshead:2009cb}. These type of large logarithmic corrections have since been ruled out on the basis of symmetry, and also correctly including different terms which should contribute to the same order \cite{Senatore:2009cf, Tanaka:2013caa}.

Thus, our result is consistent with this broad consensus that the quantum corrections calculated in an open EFT approach does not contain any large logarithmic factors (coming from the UV) which also spoils the scale invariance of the spectrum. In fact, in the late time limit, which is what one should naturally consider for evaluating cosmological correlation functions, our result shows that the leading order quantum corrections exactly cancel. There is no extra time-dependence introduced in the tensor power spectrum coming from quantum corrections due to logarithmic terms depending on the comoving momentum, which has been the main finding of loop corrections \cite{Senatore:2012ya}. 

For the benefit of the reader, let us rewrite our final result in the form:
\begin{eqnarray}\label{Main_Result}
	\left\langle h_{ij} h_{ij}\right\rangle \sim \frac{C}{k^3}\, \left(\frac{H}{\Mp}\right)^4\,\left[\ln\left(\dfrac{H}{\mu}\right) + C_2\right]\,,
\end{eqnarray}
with a negative constant $C$ and some $\mathcal{O}(1)$ constant $C_2$. Similar logarithmic terms have also appeared for loop corrections to the tensor power spectrum when considering massless isocurvature fields coupled to tensor modes, resulting in loops arising from Dirac fermions, minimally coupled scalars or Gauge fields \cite{Tan:2019czo}, or for loop corrections to the tensor power spectrum due to interactions with a conformal scalar field \cite{Frob:2012ui}. Moreover, as seen in \cite{Frob:2012ui}, there can be additive constants appearing in the one-loop correction, in addition to the $\ln(H/\mu)$ term, which depends on the renormalziation scheme. This is exactly what we find in our case as well. Another important similarity is that the prefactor $C$ of the logarithmic term is negative for all the different loop corrections considered in \cite{Frob:2012ui, Tan:2019czo}, and so is the case for us. Thus, it shows that our results are indeed along the expected lines of those coming from one loop corrections to the tensor power spectrum.

Having said that, keep in mind that the interaction terms for the aforementioned systems are determined by the coupling of the graviton to these matter fields, and not by pure GR alone (see {\it e.g.} \cite{Donoghue:1994dn} for a discussion on quantum corrections in GR as an EFT). This is the one of the  novelties of our work. The cubic interaction of the tensor modes, which leads to the quantum corrections to the power spectrum, is completely specified by the nonlinearity of GR and is free from choices of the coupling. In fact, this is why our leading terms in the interaction Hamiltonian can be free of both slow-roll parameters as well as any derivative couplings. Since we were unable to find the corresponding calculation for loop corrections to the tensor power spectrum, arising from cubic tensor interactions alone, it is impossible to do an exact comparison of our computation with standard loop corrections evaluated in the in-in formalism.  Nevertheless, this highlights a salient feature of our work: Even without going into the merits of abandoning standard loop corrections in favor of using an open EFT approach to compute quantum corrections to inflationary n-point functions, which is the main purpose of this work, we have managed to fill the lacuna in the existing literature by calculating quantum corrections to the gravitational wave  spectrum due to tensor interactions alone.

\subsection{Looking ahead: Relaxing the Markovian approximation}
As is often done when exploring new techniques, we have verified that our result matches with standard loop corrections to the tensor power spectrum and have the same logarithmic dependence on the Hubble parameter. Although this is a nice sanity check to make sure that our computations are correct to the leading order, the whole point of undertaking this exercise is to be able to go beyond loop corrections which necessarily have Wilsonian EFT underpinning its technical implementation. As we have emphasized several times, the late-time limit of the first-order quantum correction calculated here is under the strict Markovian approximation which leads to our master equation having the standard Lindblad form. Nevertheless, this is an assumption baked into our analyses -- one which we find to be self-consistent (see Appendix \ref{ap:DK}).

However, this does not imply that this assumption must necessarily be true for cosmological dynamics of the form considered here. Counterexamples of cosmological setups displaying non-Markovian behaviour have been recently demonstrated in \cite{Shandera:2017qkg, Hsiang:2021lxp}. Therefore, our goal is to next consider a more general master equation for the superhorizon modes that is time-nonlocal (of the Nakajima-Zwanzig form) \cite{nakajima1958, zwanzig1960, BreuerHeinz}. In fact, this can be seen from our calculation as well. One finds that the dissipation kernel is sharply peaked for our model (Appendix \ref{tpi}) insofar that it goes as $1/(\tau - \tau')^6$ (see Eqn. \eqref{Markovian_kernel}). Nevertheless, exact Markovian behavious would demand a delta-function peakedness which obviously is not the case for us. In other words, this shows how a systematic study can now be undertaken to explore the deviation from Markovianity for our system by allowing for more general master equations.

Therefore, the general trend seems to suggest that the relevant question to ask would be the following:  What is the regime in which we are interested in calculating quantum correction to the power spectrum due to cubic (and other higher-point) interactions? This is so because in certain regimes, the Nakajima-Zwanzig master equation does indeed reduce to the standard Markovian-Lindblad form, and if it so happens for the observational questions of interest, then we can conclude that our results are robust and our Markovian approximation justified. Indeed, this would imply that standard loop corrections implemented to calculate radiative corrections to the power spectrum of inflationary perturbations do capture the essential physics of these systems. However, this is yet to be proven and a systematic study of these systems have to be be undertaken in order to either prove the above statement or falsify it.

But the story does not end here. Apart from observational consequences, there are other conceptual lessons to be learnt from applying open EFT techniques to calculating such quantum corrections to the inflationary power spectra. Let us return to the entanglement of UV-modes to the IR ones. It might indeed be possible that the trans-Planckian modes are decoupled from observational dofs, even allowing for the non-Wilsonian character of the system.\footnote{Again, we emphasize that we have only found that absence of such entanglement effects in the strict Markovian limit and whether relaxing this approximation leads to corrections which are small or not remains to be seen.} However, if it is found that non-Markovian effects are small for observable modes and can yet become large for another subset of modes of the system, these would lead to deep ramifications for inflation (\textit{e.g.} for eternal inflation) \cite{Upper_limit} and even for QFT in dS spacetimes (\textit{e.g.} for the lifetime of metastable dS) \cite{Dvali:2018jhn, Brahma:2020tak, Brahma:2020htg}. The intriguing aspects of  non-Markovian nature of gravitational interactions are yet to explored at all and we have simply set up the mechanism to systematically investigating this question in more detail in the future.

\bigskip

\section*{Acknowledgements:}
The authors thank Thomas Colas for feedback on an earlier version of this draft. \\
SB is supported in part by the Higgs Fellowship. AB is partially supported by STFC. JCF is supported by the Secretary of Higher Education, Science, Technology and Innovation of Ecuador (SENESCYT).

\bigskip

\appendix

\section{The master equation in the Lindblad form} \label{Ame}

\subsection{Notation}

Before deriving the master equation, let us introduce some notation to avoid confusion. First, unless otherwise stated (through a subscript), every operator is written in the Schr\"odinger picture. Likewise, the subscripts $\e$ and $\s$ indicate that operators act on the environment or system Hilbert space, respectively. If no subscript is pointed out, the operator acts on the entire Hilbert space. Notice that for any operator like that we have
\begin{equation}
    \hat{A} = \sum_i \hat{A}_{\e}^i \otimes \hat{A}_{\s}^i\;,
\end{equation}
where the sum is over any possible combination of system and environment modes associated with the Fourier expansion of the operator. Naturally, this expansion is quite lengthy, but the resulting expressions can be reduced by relabeling momenta and using other symmetries, similarly to the process followed to arrive to Eqn. (\ref{HI}).  

\subsection{Building the Master Equation}

In this section, we sketch the derivation of the master equation in terms of the so-called Lindblad operators. For this, we will work in the Schr\"odinger picture, although it is convenient to start from the von-Neumann equation in the interaction picture,
\begin{equation}
	\rho_I' = -i \left[V_I (\tau),\rho_I (\tau)\right],
\end{equation} 
where $\rho_I$ is the density matrix in the interaction picture. Next, we will re-write the equation above in a way more suitable for the upcoming approximations. For this, notice that the solution is given by
\begin{equation}\label{mesol}
    \rho_I (\tau) = \rho_I (\tau_0) - i \int_{\tau_0}^{\tau} d\tau' \left[V_I (\tau'), \rho_I (\tau') \right]\,.
\end{equation}
The equation above is the full solution to the von-Neumann equation. We intend to go to second-order approximation, for which we have
\begin{equation}\label{sol2ndo}
    \rho_I (\tau) \approx \rho_I(\tau_0) - i \int_{\tau_0}^{\tau} d\tau' \left[ V_I (\tau'), \rho_I (\tau_0) \right] + (-i)^2 \int_{\tau_0}^{\tau} d\tau' \int_{\tau_0}^{\tau'} d\tau'' \left[ V_I (\tau'), \left[ V_I (\tau''), \rho_I (\tau_0) \right] \right]\,,
\end{equation}

Now, it is turn to go to the Schr\"odinger picture. To do so, we use $\rho(\tau) = U_0 (\tau;\tau_0) \rho_I (\tau) U_0^{\dagger}(\tau;\tau_0)$, so that
\begin{align}\label{expansion}
    \frac{d\rho(\tau)}{d\tau} = & -i H_0 U_0 \rho (\tau_0) U_0^{\dagger} - i (-i H_0) U_0 \int_{\tau_0}^{\tau} d\tau' \left[ V_I (\tau'), \rho_I (\tau_0) \right] U_0^{\dagger} - i U_0 \left[ V_I (\tau), \rho_I (\tau_0) \right] U_0^{\dagger} \nonumber \\
    & -i U_0 \int_{\tau_0}^{\tau} d\tau' \left[ V_I (\tau'), \rho (\tau_0) \right] U_0^{\dagger} (i H_0) + (-i)^2 (-i H_0) U_0 \int_{\tau_0}^{\tau} d\tau' \int_{\tau_0}^{\tau'} d\tau'' \left[ V_I (\tau'), \left[ V_I (\tau''), \rho (\tau_0) \right] \right] U_0^{\dagger} \nonumber \\
    & + (-i)^2 U_0 \int_{\tau_0}^{\tau} d\tau' \int_{\tau_0}^{\tau'} d\tau'' \left[ V_I (\tau'), \left[ V_I (\tau''), \rho (\tau_0) \right] \right] U_0^{\dagger} (i H_0) \nonumber \\
    & + (-i)^2 U_0 \int_{\tau_0}^{\tau} d\tau' \left[ V_I (\tau), \left[ V_I (\tau'), \rho(\tau_0) \right] \right] U_0^{\dagger}\,,
\end{align}
where we have taken the time-derivative of \eqref{sol2ndo}.

Next, in order to find the master equation governing the evolution of the reduced density matrix, we trace over the environment dofs as follows
\begin{equation}\label{tr1}
    \frac{d\rho_{\rm red}(\tau)}{d\tau} := \sum_i \left\langle {\cal E}_i \left| \frac{d\rho (\tau)}{d\tau} \right| {\cal E}_i \right\rangle\,,
\end{equation}
where $\ket{\e_i}$ denote the environment states, in our case the sub-Hubble modes at time $\tau$. Performing this operation on both sides of \eqref{expansion} we obtain
\begin{align}\label{meq}
    \frac{d \rho_{\rm red}(\tau)}{d\tau} = & -i \left[H_{0,\s}, \rho_{\rm red}^{(0)} + \rho_{\rm red}^{(1)} + \rho_{\rm red}^{(2)} \right] -i \left[V_{\rm eff1} + V_{\rm eff2}, \rho_{\s}(\tau) \right] \nonumber \\
    & - \frac{1}{2} \sum_i \left[ L_1^{\dagger}L_2 \rho_\s (\tau) + \rho_\s (\tau) L_2^{\dagger} L_1 - 2 L_1 \rho_\s (\tau) L_2^{\dagger} + (L_1 \leftrightarrow L_2 ) \right]\,,
\end{align}
where
\begin{equation}
    \rho_I (\tau_0) = \rho(\tau_0) = \ket{\e_0}\ket{\s_0}\bra{\s_0}\bra{\e_0}\,,
\end{equation}
\begin{equation}
    \rho_{\rm red}^{(0)} (\tau) := \rho_\s (\tau) = U_{0,\s}\ket{\s_0}\bra{\s_0}U_{0,\s}^{\dagger} = \ket{\s(\tau)}\bra{\s(\tau)}\,,
\end{equation}
\begin{equation}
    V_{\rm eff1} := \left\langle \e (\tau_0) \left| U_{0,\e}^{\dagger} (\tau; \tau_0) V_{\s} U_{0, \e}(\tau;\tau_0) \right| \e (\tau_0) \right\rangle\,, \quad V_{\rm eff2} := -\frac{i}{2} \sum_i \left(L_1^{\dagger}L_2 - L_2^{\dagger}L_1\right)\,,
\end{equation}
\begin{equation}
    L_1 := \Big\langle \e_i \Big| V_{\s} U_{0,\e}(\tau;\tau_0)\Big| \e (\tau_0) \Big\rangle\,, \quad    L_2 := \left\langle \e_i \left| \int_{\tau_0}^{\tau} d\tau' V_I (\tau' - \tau) U_{0,\e}(\tau;\tau_0) \right| \e(\tau_0) \right\rangle\,,
\end{equation}
and $V_I (\tau'-\tau) = U_0^{\dagger} (\tau';\tau) V U_0 (\tau';\tau)$\,. $L_1$ and $L_2$ are the so-called Lindblad operators. 

In order to see how Eqn. (\ref{meq}) comes about, notice that every term containing an $H_0$ factor in \eqref{expansion} contributes to the commutator of the quadratic Hamiltonian in \eqref{meq}. Regarding the effective Hamiltonians, the first one comes directly from the third term on the first line, on the r.h.s. of \eqref{expansion}. The second effective Hamiltonian comes (partially) from the last term on \eqref{expansion}. To see this, notice that 
\begin{align}
    & \sum_i (-i)^2  \left\langle \e_i \left|  U_0(\tau;\tau_0) \int_{\tau_0}^{\tau} d\tau' \left[V_I (\tau), \left[ V_I(\tau'), \rho (\tau_0) \right] \right] U_0^{\dagger} (\tau;\tau_0) \right| \e_i \right\rangle \nonumber \\
    & = (-i)^2 \sum_i \left\{ L_1^{\dagger}L_2 \rho_\s (\tau) + \rho_\s (\tau) L_2^{\dagger}L_1 - L_2 \rho_\s (\tau) L_1^{\dagger} - L_1 \rho_\s (\tau) L_2^{\dagger} \right\} \nonumber \\
    & = -i \left[H_{\rm int}^{\rm(eff2)}, \rho_\s (\tau)\right] - \frac{1}{2} \sum_i \left\{L_1^{\dagger}L_2 \rho_\s (\tau) + \rho_\s (\tau) L_2^{\dagger}L_1 - 2 L_1 \rho_\s (\tau) L_2^{\dagger} + (L_1 \leftrightarrow L_2) \right\}\,.
\end{align}
It can be seen that in order to construct $V_{\rm eff2}$ we summed and subtracted $\frac{1}{2} \sum_i \left[ L_2^{\dagger} L_1, \rho_\s (\tau) \right]$ to the expression,  which renders the formula above. In this way, the last term encompasses the non-unitary part of the evolution, as the remaining terms can be written as a commutator with a Hamiltonian, which can be seen as unitary evolution.

\section{Two–point function integrals}\label{tpi}

The correction to the power spectrum is determined by Eqn. \eqref{corx}, which we reproduce below for the convenience of the reader:
\begin{align}\label{corx1}
    & \Tr \left[ \hat{\cal O}_{\vt q} \rho_{\rm red}^{(2)} (\tau) \right] = - \frac{2}{a^2 \Mp^2} \times 18 \int_{-1/q}^{\tau} d\tau' \int_{\e} \frac{d^3 k_2}{(2\pi)^3}\int_{-1/q}^{\tau'} d\tau''  \bigg\{ e_{ij}^{\alpha} ({\vt q}) e_{ji}^{\alpha} (- {\vt q}) \left( v_q (\tau) \right)^2 \nonumber \\   
    & \bigg[ h_1 (\tau', -{\vt q}, {\vt k_2}, - {\vt q} - {\vt k_2}) h_0^* (\tau'', \vt{q}, {\vt k_2}, -{\vt q} - {\vt k_2}) 
     + h_1 (\tau', {\vt q}, {\vt k_2}, {\vt q} - {\vt k_2}) h_0^* (\tau'', -\vt{q}, {\vt k_2}, {\vt q} - {\vt k_2}) \bigg] \delta_{\alpha, \alpha_1}\delta_{\alpha,\beta_1} \nonumber \\
    & - \bigg[ e_{ij}^{\alpha} ({\vt q}) e_{ji}^{\alpha} ({\vt q}) h_0 (\tau', {\vt q}, {\vt k_2}, -{\vt q} - {\vt k_2}) h_0^* (\tau'', {\vt q}, {\vt k_2}, -{\vt q} - {\vt k_2}) \nonumber \\
    & + e_{ij}^{\alpha} (- \vt{q}) e_{ji}^{\alpha} (-\vt{q}) h_0(\tau', -{\vt q}, {\vt k_2}, {\vt q} - {\vt k_2}) h_0^* (\tau'', -{\vt q}, {\vt k_2}, {\vt q} - {\vt k_2}) \bigg] \left| v_q (\tau) \right|^2 \delta_{\alpha, \alpha_1}\delta_{\alpha, \beta_1} + {\rm c.c.} \bigg\} \;.
\end{align}
Then, upon the introduction of the dimensionless variables $\omega'$ and $\kappa_i$, the integrals we need to solve are of the type
\begin{align}
  I_{1,i} := \int_{1}^{\omega} \frac{d\omega'}{(\omega')^2} \int_{1}^{\infty} d\kappa_2  \int_{-1}^{1} dy\ \frac{\kappa_2}{\kappa_3} f_i(y) \beta(\omega')\beta(\kappa_2) \beta(\kappa_3) \Xi (\omega', \kappa_2, y)\;,
\end{align}
\begin{align}
  I_{2,i} := \int_{1}^{\omega} \frac{d\omega'}{(\omega')^2} \int_{1}^{\infty} d\kappa_2  \int_{-1}^{1} dy\ \frac{\kappa_2}{\kappa_3} f_i(y) \beta^*(\omega')\beta(\kappa_2) \beta(\kappa_3) \Xi (\omega', \kappa_2, y)\;,
\end{align}
where $i$ is an index referring to the two functions $f_1(\theta)$ and $f_2 (\theta)$ –Eqs. \eqref{eqf1} and \eqref{eqf2}–, which we found by summing over the polarisation tensors present in Eqn. \eqref{corx}.

Next, as explained in the main text, we expand the integrand around $\kappa_2 \rightarrow \infty$, which is the region that dominates the integral. We have checked that the same approximation gives a good order–of–magnitude approximation for the region near $\kappa_2 = 1$. However, and rather conveniently, the contributions from that limit cancel out, leaving us with the UV-divergent terms.

With these considerations, we show below the resulting expressions from each term in Eqn. \eqref{corx1}, including their complex conjugates. 

\begin{align}\label{UV1}
  \text{1st term:}\; & - \frac{9}{2\pi^5} \frac{H^4}{\Mp^4} \omega^2 \left[e^{i\omega} (1+i/\omega)\right]^2 I_{2,1} + \text{c.c.} \nonumber \\
   = & - \frac{9}{2\pi^5} \frac{H^4}{\Mp^4} \Bigg\{ \frac{128\pi}{45} \left[-2 + e^{2i(-1+\omega)} \Big[(1-i)+e^{2i}\left({\rm Ei}(-2i) - {\rm Ei}(-2i\omega)\right)\Big] + {\rm c.c.} \right] \ln \frac{\Lambda}{H} \nonumber \\
   & + \frac{128\pi}{45} \bigg[-1 + \frac{1}{\omega^3} + \frac{3}{2\omega^2} (-1 + 30 \pi (1 + 5 \cos 2 + \sin 2)) + \frac{1}{\omega^3}\bigg]\Bigg\}\;,
\end{align}
\begin{align}\label{UV2}
  \text{2nd term:}\; & - \frac{9}{2\pi^5} \frac{H^4}{\Mp^4} \omega^2 \left[e^{i\omega} (1+i/\omega)\right]^2 I_{2,2} + \text{c.c.} \nonumber \\
   = & - \frac{9}{2\pi^5} \frac{H^4}{\Mp^4} \Bigg\{ \frac{128\pi}{45} \left[-2 + e^{2i(-1+\omega)} \Big[(1-i)+e^{2i}\left(({\rm Ei}(-2i) - {\rm Ei}(-2i\omega)\right)\Big] + {\rm c.c.} \right] \ln \frac{\Lambda}{H} \nonumber \\
   & + \frac{128\pi}{45} \bigg[-1 + \frac{1}{\omega^3} + \frac{3}{2\omega^2} (-1 + 30 \pi (1 + 5 \cos 2 + \sin 2)) + \frac{1}{\omega^3}\bigg]\Bigg\}\;,
\end{align}
\begin{align}\label{Tame1}
  \text{3rd term:}\; & \frac{9}{2\pi^5} \frac{H^4}{\Mp^4} \omega^2 \left|e^{i\omega} (1+i/\omega)\right|^2 I_{1,1} + \text{c.c.} \nonumber \\
   = & \frac{9}{2\pi^5} \frac{H^4}{\Mp^4} \Bigg\{ \frac{128\pi}{45} \bigg[-1 + \frac{1}{\omega^3} + \frac{3}{2\omega^2} (-1 + 30 \pi (1 + 5 \cos 2 + \sin 2)) + \frac{1}{\omega^3}\bigg]\Bigg\}\;,
\end{align}
\begin{align}\label{Tame2}
  \text{4th term:}\; & \frac{9}{2\pi^5} \frac{H^4}{\Mp^4} \omega^2 \left|e^{i\omega} (1+i/\omega)\right|^2 I_{1,2} + \text{c.c.} \nonumber \\
   = & \frac{9}{2\pi^5} \frac{H^4}{\Mp^4} \Bigg\{ \frac{128\pi}{45} \bigg[-1 + \frac{1}{\omega^3} + \frac{3}{2\omega^2} (-1 + 30 \pi (1 + 5 \cos 2 + \sin 2)) + \frac{1}{\omega^3}\bigg]\Bigg\}\;.
\end{align}
Notice that each expression above represents the contribution to the dimensionless tensor power spectrum, which is why they are dimensionless. 

The behavior of the function multiplying the UV–divergent term might appear rather obscure due to the Ei functions. To get a better grasp of this function near $\omega \rightarrow 0$, we Taylor expand it around that value, obtaining 
\begin{equation}
    {\rm Ei}(2i \omega) \approx \gamma_E + \frac{i\pi}{2} + \ln (2\omega) + {\cal O}(\omega)\;.
\end{equation}
Then, near $\omega = 0$, {\it i.e.,} in the late–time limit, the first and second terms have the form
\begin{align}
  \text{1st \& 2nd terms:}\; & - \frac{9}{2\pi^5} \frac{H^4}{\Mp^4} \Bigg\{ \frac{256\pi}{45} \left[-2 + \gamma_E - \cos 2 + \sin 2 - {\rm Ci}\ 2 + \ln(2\omega) \right] \ln \frac{\Lambda}{H} \nonumber \\
   & + \frac{128\pi}{45} \bigg[-1 + \frac{1}{\omega^3} + \frac{3}{2\omega^2} (-1 + 30 \pi (1 + 5 \cos 2 + \sin 2)) + \frac{1}{\omega^3}\bigg]\Bigg\}\;,
\end{align}
Then, adding the equations above, we easily recover Eqn. \eqref{fincon}.

\section{Dissipation Kernel}\label{ap:DK}

In principle, one can also find the master equation in terms of a dissipation kernel, similar to the process followed by \cite{Boyanovsky:2015tba, Brahma:2021mng}. Such an approach proved to be less advantageous than the one used in this work when trying to find the corrections to the spectrum, although one can still find a kernel characterizing the dynamics of the environment in its interaction with the system. To do so, consider the interaction of the form
\begin{equation}
	\hat{V}_I (\tau) = -2 \left(1-\frac{\epsilon}{3}\right) a^2 \Mp^2 {\cal H}^2 \int_{\Delta_k} \hat{h}_{ij}^{\s} (\tau, \vt{k}_1) \hat{h}_{jk}^{\e} (\tau, \vt{k}_2) \hat{h}_{ki}^{\e} (\tau, \vt{k}_3)\;, 
\end{equation}
where we have chosen the same combination of environment and system degrees of freedom as before. Next, we re-write the interaction Hamiltonian in terms of the Bunch-Davies mode functions, such that
\begin{equation}
	\hat{V}_I (\tau) = 4 \sqrt{2} \frac{H}{\Mp}\frac{1}{\tau} \int_{\Delta_k} \sum_{\{\alpha_i\}} [\hat{\cal V}^{\s}_{\vt{k}_1} (\tau)]_{ij}^{\alpha_1}\ [\hat{\cal V}^{\e}_{\vt{k}_2} (\tau)]_{jk}^{\alpha_2}\ [\hat{\cal V}^{\e}_{\vt{k}_3} (\tau)]_{ki}^{\alpha_3},
\end{equation}
where we have introduced the following operator for future convenience
\begin{equation}\label{BDop}
    [\hat{\cal V}^{\s}_{\vt{k}_1} (\tau)]_{ij}^{\alpha_1} := v_k (\tau) e_{ij}^{\alpha} ({\vt k}) \aop{a}{k}^{\alpha}  + v_k^* (\tau) e_{ij}^{\alpha} (-{\vt k}) \aop{a}{- \vt k}^{\alpha \dagger}\;.
\end{equation}
Finally, from here onwards, we denote the coupling as $\lambda (\tau) := 4 \sqrt{2} H/ (\tau \Mp)$. 

\subsection{Master equation}

First, we shall work in the interaction picture in contrast to the process followed in the main text, so every operator should be understood to belong to that picture. With this consideration in mind, the density matrix is given by
\begin{equation}
	\rho_I(\tau) = \rho_{\s} (\tau) \otimes \rho_{\e} (\tau_0)\;,
\end{equation}
where
\begin{equation}\label{rhoenv}
    \rho_\e (\tau_0) = \ket{\e_0} \bra{\e_0} = \ket{0} \bra{0}\;.
\end{equation}
Notice that environment and system degrees of freedom are defined exactly as before, i.e., they are delimited by the Hubble length. Furthermore, since the quadratic Hamiltonian governs the evolution of states in the interaction picture, the factorisation of the density matrix holds at later times. Moreover, the weak coupling between system and environment keeps the latter unperturbed, such that $\rho_{\e} (\tau) \approx \rho_{\e} (\tau_0)$. Then, tracing over the environment degrees of freedom on Eqn. (\ref{sol2ndo}), we get
\begin{align}\label{tsm}
	\rho_{{\rm r},I}' (\tau) &= - \lambda(\tau) \int_{\tau_0}^{\tau} \lambda(\tau') \int_{\Delta_k}\int_{\Delta_p} \sum_{\{\alpha_i,\beta_i\}}\nonumber \\
	& \Big\{ [\hat{\cal V}^{\s}_{\vt{k}_1} (\tau)]_{ij}^{\alpha_1} [\hat{\cal V}^{\s}_{\vt{p}_1} (\tau')]_{lm}^{\beta_1} \Tr_{\e} \left[ [\hat{\cal V}^{\e}_{\vt{k}_2} (\tau)]_{jk}^{\alpha_2} [\hat{\cal V}^{\e}_{\vt{k}_3} (\tau)]_{ki}^{\alpha_3} [\hat{\cal V}^{\e}_{\vt{p}_2} (\tau')]_{mn}^{\beta_2} [\hat{\cal V}^{\e}_{\vt{p}_3} (\tau')]_{nl}^{\beta_3} \rho_\e (\tau_0) \right] \rho_{{\rm r},I} (\tau') \nonumber \\
	&- [\hat{\cal V}^{\s}_{\vt{k}_1} (\tau)]_{ij}^{\alpha_1} \Tr_{\e} \left[ [\hat{\cal V}^{\e}_{\vt{k}_2} (\tau)]_{jk}^{\alpha_2} [\hat{\cal V}^{\e}_{\vt{k}_3} (\tau)]_{ki}^{\alpha_3} \rho_\e (\tau_0) [\hat{\cal V}^{\e}_{\vt{p}_2} (\tau')]_{mn}^{\beta_2} [\hat{\cal V}^{\e}_{\vt{p}_3} (\tau')]_{nl}^{\beta_3} \right] \rho_{{\rm r},I} (\tau') [\hat{\cal V}^{\s}_{\vt{p}_1} (\tau')]_{lm}^{\beta_1} \nonumber \\
	&- [\hat{\cal V}^{\s}_{\vt{p}_1} (\tau')]_{lm}^{\beta_1} \Tr_{\e} \left[ [\hat{\cal V}^{\e}_{\vt{p}_2} (\tau')]_{mn}^{\beta_2} [\hat{\cal V}^{\e}_{\vt{p}_3} (\tau')]_{nl}^{\beta_3} \rho_\e (\tau_0) [\hat{\cal V}^{\e}_{\vt{k}_2} (\tau)]_{jk}^{\alpha_2} [\hat{\cal V}^{\e}_{\vt{k}_3} (\tau)]_{ki}^{\alpha_3} \right] \rho_{{\rm r},I} (\tau') [\hat{\cal V}^{\s}_{\vt{k}_1} (\tau)]_{ij}^{\alpha_1} \nonumber \\
	&+ \rho_{{\rm r},I}(\tau')  [\hat{\cal V}^{\s}_{\vt{p}_1} (\tau')]_{lm}^{\beta_1}  \Tr_\e \left[\rho_\e (\tau_0) [\hat{\cal V}^{\e}_{\vt{p}_2} (\tau')]_{mn}^{\beta_2} [\hat{\cal V}^{\e}_{\vt{p}_3} (\tau')]_{nl}^{\beta_3} [\hat{\cal V}^{\e}_{\vt{k}_2} (\tau)]_{jk}^{\alpha_2} [\hat{\cal V}^{\e}_{\vt{k}_3} (\tau)]_{ki}^{\alpha_3} \right]  [\hat{\cal V}^{\s}_{\vt{k}_1} (\tau)]_{ij}^{\alpha_1}\Big\}\;,
\end{align}
where $\rho_{{\rm r},I}$ denotes the reduced density matrix in the interaction picture. We can readily compute the traces by using eqs. (\ref{BDop}) and (\ref{rhoenv}), which yields inner products on the Hilbert space of the environment degrees of freedom as follows 
\begin{equation}
\left\langle 1_{\vt{k}_2}^{\alpha_2}, 1_{\vt{k}_3}^{\alpha_3} \Big| 1_{-\vt{p}_2}^{\beta_2}, 1_{-\vt{p}_3}^{\beta_3} \right\rangle = \delta_{\alpha_2,\beta_2}\delta_{\alpha_3,\beta_3}\delta(\vt{k}_2+\vt{p}_2)\delta(\vt{k}_3+\vt{p}_3) + \delta_{\alpha_2,\beta_3}\delta_{\alpha_3,\beta_2}\delta(\vt{k}_2+\vt{p}_3) \delta(\vt{k}_3+\vt{p}_2) .
\end{equation}
This product forces the same polarisation on the tensors, as well as the momenta. Then, one can define a kernel matrix by integrating over one of the environment modes --$k_2$ in this case-- and remembering that the other one, $k_3$, is fixed by the Dirac delta. Then, the elements of this matrix are of the form
\begin{equation}
 {\cal K}_{jiml}^{\alpha_2\alpha_3} \sim \int \frac{d^3 k_2}{(2\pi)^3} v_{k_2} (\tau) v_{k_2}^* (\tau') v_{k_3} (\tau) v_{k_3}^* (\tau')  e^{\alpha_2}_{jk}(\vt{k}_2) e^{\alpha_3}_{ki}(\vt{k}_3) \left[e^{\alpha_2}_{mn}(\vt{k}_2)e^{\alpha_3}_{nl}(\vt{k}_3) + e^{\alpha_3}_{mn}(\vt{k}_3)e^{\alpha_2}_{nl}(\vt{k}_2)  \right]\;,  
\end{equation}
with the corresponding term in the master equation reading
\begin{equation}
	\rho_{{\rm r},I}'(\tau) \sim -\lambda(\tau) \int_{\tau_0}^{\tau} d\tau' \lambda(\tau') \sum_{\vt{k}_1} \sum_{\{\alpha_i\},\beta_1} \Big\{ [\hat{\cal V}^{\s}_{\vt{k}_1} (\tau)]_{ij}^{\alpha_1} [\hat{\cal V}^{\s}_{-\vt{k}_1} (\tau')]_{lm}^{\beta_1} {\cal K}_{jiml}^{\alpha_2\alpha_3} (\vt{k}_1,\tau) + \cdots
\end{equation}
\subsection{Computing the kernel}

We proceed by computing the sum over the polarisation states, which will allow us to define a proper kernel akin to that for scalar perturbations worked out in \cite{Berera:2020dvn}. For this, we need to perform the integral over one of the sub-horizon momenta, say $\vt{k}_2$, which forces us to work with explicit expressions for the polarisation tensors corresponding to $\vt{k}_1$, $\vt{k}_2$ and $\vt{k}_3$. We proceed similarly as in Section \ref{PCPP}, i.e., we align the system mode to the $\vt{z}-$direction so that
\begin{equation*}
\vt{k}_1 = k_1(0,0,1), \vt{k}_2 = k_2(\sin \theta \cos \phi, \sin \theta \sin \phi, \cos \theta),  \vt{k}_3 = -(k_2 \sin \theta \cos \phi, k_2 \sin \theta \sin \phi, k_1 + k_2 \cos \theta),
\end{equation*}
where, for the last equation, we have used the fact that $\vt{k}_3 = - (\vt{k}_1 + \vt{k}_2)$. With these considerations, the polarisation tensors are
\begin{equation}\label{polten1}
	e^{+} (\hat{\vt{k}}_1) = \left( \begin{array}{ccc} 1 & 0 & 0 \\ 0 & -1 & 0 \\ 0 & 0 &0 \end{array} \right), \qquad e^{\times} (\hat{\vt{k}}_1) = \left( \begin{array}{ccc} 0 & 1 & 0 \\ 1 & 0 & 0 \\ 0 & 0 &0 \end{array} \right),
\end{equation}
\begin{equation*}
 e^{+} (\hat{\vt{k}}_2) = \left( \begin{array}{ccc} \cos^2 \theta \cos^2 \phi - \sin^2 \phi & (1 + \cos^2 \theta) \sin \phi \cos \phi & - \sin \theta \cos \theta \cos \phi \\
 (1+\cos^2 \theta) \sin \phi \cos \phi & \cos^2 \theta \sin^2 \phi - \cos^2 \theta & - \sin \theta \cos \theta \sin \phi \\
 -\sin \theta \cos \theta \cos \phi & - \sin \theta \cos \theta \sin \phi & \sin^2 \theta
 \end{array} \right),
\end{equation*}
\begin{equation}\label{polten2}
 e^{\times} (\hat{\vt{k}}_2) = \left( \begin{array}{ccc} -2 \cos \theta \sin \phi \cos \phi & \cos \theta (\cos^2 \phi - \sin^2 \phi) & \sin \theta \sin \phi \\
 \cos \theta (\cos^2 \phi - \sin^2 \phi) & 2 \cos \theta \sin \phi \cos \phi & -\sin \theta \cos \phi \\
 \sin \theta \sin \phi & -\sin \theta \cos \phi & 0 \end{array} \right),
 \end{equation} 
 \begin{equation*}
e^{+} (\hat{\vt{k}}_3)  = \left(\begin{array}{ccc}
\frac{(k_1+k_2 \cos  \theta )^{2} \cos^2  \phi}{k_1^{2}+k_2^{2}+2 k_1 k_2 \cos  \theta }-\sin^2  \phi & \left(1+\frac{(k_1+k_2 \cos  \theta )^{2}}{k_1^{2}+k_2^{2}+2 k_1 k_2 \cos  \theta }\right) \cos  \phi  \sin  \phi  & -\frac{k_2(k_1+k_2 \cos  \theta ) \cos  \phi  \sin  \theta }{k_1^{2}+k_2^{2}+2 k_1 k_2 \cos  \theta } \\

\left(1+\frac{(k_1+k_2 \cos  \theta )^{2}}{k_1^{2}+k_2^{2}+2 k_1 k_2 \cos  \theta }\right) \cos  \phi  \sin  \phi  & -\cos^2  \phi+\frac{(k_1+k_2 \cos  \theta )^{2} \sin^2  \phi}{k_1^{2}+k_2^{2}+2 k_1 k_2 \cos  \theta } & -\frac{k_2(k_1+k_2 \cos  \theta ) \sin  \theta  \sin  \phi }{k_1^{2}+k_2^{2}+2 k_1 k_2 \cos  \theta } \\

-\frac{k_2(k_1+k_2 \cos  \theta ) \cos  \phi  \sin  \theta }{k_1^{2}+k_2^{2}+2 k_1 k_2 \cos  \theta } & -\frac{k_2(k_1+k_2 \cos  \theta ) \sin  \theta  \sin  \phi }{k_1^{2}+k_2^{2}+2 k_1 k_2 \cos  \theta } & \frac{k_2^{2} \sin^2  \theta}{k_1^{2}+k_2^{2}+2 k_1 k_2 \cos \theta }
\end{array}\right),
 \end{equation*}
\begin{equation}\label{polten3}
e^{\times} (\hat{\vt{k}}_3) =  \left(\begin{array}{cccc}
\frac{(k_1+k_2 \cos \theta) \sin 2 \phi}{\sqrt{k_1^{2}+k_2^{2}+2 k_1 k_2 \cos \theta}} & -\frac{(k_1+k_2 \cos \theta) \cos 2 \phi}{\sqrt{k_1^{2}+k_2^{2}+2 k_1 k_2 \cos \theta}} & -\frac{k_2 \sin \theta \sin \phi}{\sqrt{k_1^{2}+k_2^{2}+2 k_1 k_2 \cos \theta}} \\
-\frac{(k_1+k_2 \cos \theta) \cos 2 \phi}{\sqrt{k_1^{2}+k_2^{2}+2 k_1 k_2 \cos \theta}} & -\frac{(k_1+k_2 \cos \theta) \sin 2 \phi}{\sqrt{k_1^{2}+k_2^{2}+2 k_1 k_2 \cos \theta}} & \frac{k_2 \cos \phi \sin \theta}{\sqrt{k_1^{2}+k_2^{2}+2 k_1 k_2 \cos \theta}} \\
-\frac{k_2 \sin \theta \sin \phi}{\sqrt{k_1^{2}+k_2^{2}+2 k_1 k_2 \cos \theta}} & \frac{k_2 \cos \phi \sin \theta}{\sqrt{k_1^{2}+k_2^{2}+2 k_1 k_2 \cos \theta}} & 0
\end{array}\right).
\end{equation}
 
Next, computing the sum over repeated indices and integrating over the azimuthal angle, we have that
 \begin{align}
	\rho_{{\rm r},I}'(\tau) \sim \lambda(\tau) \int d\tau' \lambda(\tau') \sum_{\vt{k}}& K_{k}(\tau,\tau') \Big\{ \left[ v_k (\tau) \aop{a}{k}^{+} + v_k^* (\tau) \aop{a}{-k}^{+ \dagger}\right] \left[ v_k (\tau') \aop[-]{a}{k}^{+} + v_k^* (\tau') \aop{a}{k}^{+ \dagger}\right] \nonumber \\
	&- \left[ v_k (\tau) \aop{a}{k}^{\times} - v_k^* (\tau) \aop{a}{-k}^{\times \dagger}\right] \left[ v_k (\tau') \aop{a}{-k}^{\times} - v_k^* (\tau')  \aop{a}{k}^{\times \dagger}\right]  \Big\},
\end{align}
where
\begin{equation}
	K_{k_1} (\tau,\tau') = \int_{\e} \frac{dk_2}{(2\pi)^3} k_2^2\int d(\cos \theta)\ g(k_1,k_2,\cos \theta) v_{k_2} (\tau) v_{k_2}^* (\tau') v_{k_3} (\tau) v_{k_3}^* (\tau') ,
\end{equation}
\begin{equation}
g(k_1,k_2,\cos \theta) = -\pi \frac{4k_1^4 + 11 k_1^2 k_2^2 + 4k_2^4 + 8k_1 k_2 (k_1^2 + k_2^2) \cos \theta + k_1^2 k_2^2 \cos 2\theta}{(k_1^2 + k_2^2 + 2 k_1 k_2 \cos \theta)^2}.
\end{equation}Notice the similarity of the equation above to eqs. \eqref{eqf1}–\eqref{eqf2}.
In order to compute the integral analytically, we assume a (very) squeezed configuration, such that $k_2 \sim k_3 \gg k_1$, which renders
\begin{equation}\label{Markovian_kernel}
	K_k (\tau,\tau') \approx - \frac{i e^{2i(\tau-\tau')/\tau} \left[3 k (\tau-\tau') \cos (k(\tau-\tau')) + (k^2 (\tau-\tau')^2 - 3) \sin (k(\tau-\tau'))\right]}{\pi^2 k^5 (\tau-\tau')^6}\;.
\end{equation}

As hinted before, making further progress from this point becomes rather intractable and therefore, we switch to the alternative approach detailed in the main text. However, the explicit form of the kernel depicts one of the most crucial features of the studied system, the ultra time-locality of the kernel, which is encompassed in its denominator. This implies that the time scales of the physical processes in the environment are much smaller than those of the system. Thus, this sharply-peaked nature of the kernel is to be construed as a signal of the assumed Markovian behavior \cite{BreuerHeinz, boyanovsky2017heisenberg}.

\end{document}